\documentclass[lettersize,journal]{IEEEtran}
\usepackage{amsmath,amssymb,amsfonts}
\usepackage{algorithm}
\usepackage{algpseudocode}
\usepackage{array}
\usepackage[caption=false,font=normalsize,labelfont=sf,textfont=sf]{subfig}
\usepackage{textcomp}
\usepackage{stfloats}
\usepackage{url}
\usepackage{verbatim}
\usepackage{graphicx}
\usepackage{xcolor}
\usepackage{hyperref}
\usepackage{cite}
\usepackage{array, makecell}
\usepackage{mathtools}
\usepackage{booktabs}

\usepackage[font=small,labelfont=bf]{caption}
\begin{document}

\title{An Uncertainty Quantification Framework for Deep Learning-Based Automatic Modulation Classification}

\author{Huian Yang and Rajeev Sahay

\thanks{H. Yang and R. Sahay are with the Department of Electrical and Computer Engineering, UC San Diego, San Diego, CA, 92093 USA. E-mail: \{huy011,r2sahay\}@ucsd.edu.}
\thanks{This work was supported in part by the UC San Diego Academic Senate under grant RG114404 and in part by the National Science Foundation (NSF) under grant 2512912.}}



\maketitle

\begin{abstract}

Deep learning has been shown to be highly effective for automatic modulation classification (AMC), which is a pivotal technology for next-generation cognitive communications. Yet, existing deep learning methods for AMC often lack robust mechanisms for uncertainty quantification (UQ). This limitation restricts their ability to produce accurate and reliable predictions in real-world environments, where signals can be perturbed as a result of several factors such as interference and low signal-to-noise ratios (SNR). To address this problem, we propose a deep ensemble approach that leverages multiple convolutional neural networks (CNNs) to generate predictive distributions, as opposed to point estimates produced by standard deep learning models, which produce statistical characteristics that quantify the uncertainty associated with each prediction. We validate our approach using real-world AMC data, evaluating performance through multiple UQ metrics in a variety of signal environments. Our results show that our proposed ensemble-based framework captures uncertainty to a greater degree compared to previously proposed baselines in multiple settings, including in-distribution samples, out-of-distribution samples, and low SNR signals. These findings highlight the strong UQ capabilities of our ensemble-based AMC approach, paving the way for more robust deep learning-based AMC.

\end{abstract}


\section{Introduction}

\IEEEPARstart{A}{utomatic} Modulation Classification (AMC) is a vital component of modern wireless networks, as it enables the identification of modulation schemes without prior knowledge of the transmitted signal. This capability is crucial for ensuring reliable communication in dynamic and noisy environments, including military \cite{mil_apps1,mil_apps2} and civilian applications \cite{ref6}, where signals can be subject to interference and environmental distortions. As the Internet of Things (IoT) expands, next-generation communications (e.g., 6G) will heavily rely on AMC for efficient management of the increasingly congested wireless spectrum \cite{ref50,amc_survey}. 

Traditional AMC methods, such as maximum likelihood-based (MLB) approaches \cite{ref3, ref33, ref34, ref35}, rely on statistical tests to estimate the probability of a particular modulation scheme. Although these methods can be effective, they require substantial prior knowledge 
of the received signal and its channel conditions \cite{ref29}. Furthermore, MLB methods tend to be computationally expensive, incurring high latencies while requiring manual feature engineering from received in-phase and quadrature (IQ) time samples. Such limitations restrict their scalability and adaptability into the high-volume IoT spectrum in real-world scenarios.

To address these limitations, deep learning (DL) has recently emerged as a powerful alternative to MLB methods for AMC. DL models can efficiently learn distinguishing AMC characteristics from received IQ signals without requiring manual feature engineering while achieving state-of-the-art classification accuracy \cite{ref32, ref6, ref24, ref25, ref49}. As a result, they provide a promising alternative to MLB approaches to meet the demands of overcrowded next-generation cellular networks. However, despite their success, DL models exhibit a critical limitation. Specifically, deep learning-based AMC systems are frequently \emph{overconfident} \cite{ref8} in their predictions, even when they are incorrect, and often lack robust mechanisms for uncertainty quantification (UQ) to characterize the confidence associated with a given prediction. This overconfidence can result in significant performance degradations during deployment, particularly on incorrect predictions, when the model encounters previously unseen signal environments. Such overconfidence is further exacerbated in mission-critical settings, where decisions must be made under uncertainty, including low SNR conditions and adversarial scenarios \cite{ref53, ref55}.

To address these challenges, we propose a UQ framework for AMC based on an ensemble of convolutional neural networks (CNNs), which generate predictive distributions, as opposed to point estimates produced by current state-of-the-art CNNs, whose statistical characteristics quantify the uncertainty associated with each prediction. Building on ensemble learning techniques \cite{ref11, ref12, ref15, ref18, ref19}, our approach provides calibrated uncertainty estimates, thereby enhancing reliability and robustness of current DL-based AMC frameworks. Compared to traditional UQ methods such as Bayesian Neural Networks (BNNs) \cite{ref36}, our proposed framework offers superior performance with lower computational overhead. Specifically, our method achieves higher classification accuracy in multiple signal environments, demonstrates better performance on multiple UQ metrics, and displays enhanced resilience on out of distribution (OOD) samples such as adversarial examples \cite{amc_adv_atk, amc_adv_atk2, ref40}. Moreover, our ensemble-based approach is scalable to large-parameter architectures such as ResNet \cite{ref37}, enabling existing DL-based AMC frameworks to improve their performance while simultaneously incorporating UQ for increased robustness.

To our knowledge, this is the first work to evaluate the efficacy of deep ensembles for AMC. While ensemble techniques have been used in deep learning, their application in AMC poses unique challenges due to domain-specific distortions (e.g., dynamic SNR \cite{ref32} and adversarial AMC conditions \cite{amc_adv_atk}), and our results demonstrate that equal-weighted ensembles outperform more complex methods like BNNs and SNR-aware weighted ensembles in both classification and uncertainty estimation.

Our specific contributions of this work can be summarized as follows:
\begin{enumerate}
    \item \textbf{Deep Ensemble Framework for AMC}: We develop a robust ensemble-based AMC framework comprised of multiple CNNs. \emph{To the best of our knowledge, this is the first ensemble-based deep learning UQ framework for AMC.} 
    \item \textbf{UQ Characterization in AMC}:  For the first time, we develop UQ metrics that characterize the uncertainty associated with each received signal's modulation prediction. In comparison to other considered baselines, under a range of UQ metrics, we show that our method demonstrates significantly higher robustness by consistently scoring higher on these metrics.
    \item \textbf{Out of Distribution Testing}: We test our framework on OOD samples, via adversarial perturbations, and show improved resistance to these perturbations, highlighting its UQ capability on shifted signal distributions. 
\end{enumerate}

The remainder of this paper is organized as follows. Sec. \ref{re_wrks} reviews related work on DL-based AMC including uncertainty estimation in AMC, the robustness of ensemble learning, and recent expansions of UQ in AMC. Sec. \ref{methods} outlines our methodology, detailing the signal modeling process, ensemble modeling, estimation metrics, and the characterization of OOD samples. Sec. \ref{sec:performance_evaluation} describes our experimental setup, presents UQ scoring and estimation results, and evaluates OOD performance. Finally, we close with concluding thoughts and future directions in Sec. \ref{conclusion_sec}. 

\section{Related Works} \label{re_wrks}

Earlier methods for AMC relied on likelihood methods \cite{ref29,ref33,ref35}, which provided theoretical guarantees as well as uncertainty estimation. However, such methods incur computational inefficiency in large-scale networks, making them difficult to adopt in the increasingly congested wireless spectrum. DL, as an alternative to MLB approaches, has gained significant attention for AMC due to its state-of-the-art classification performance and significantly lower inference latencies compared to MLB approaches. In this regard, various neural network architectures have been explored such as CNNs \cite{ref5,ref6,ref50,ref9}, recurrent neural networks (RNNs) \cite{ref42, ref43, ref44}, and long short-term memory (LSTM) \cite{ref41}. By learning hierarchical features directly from raw signals, these DL approaches, in addition to having low online latencies, often excel in AMC classification tasks with little to no need for feature engineering. However, while prior studies have demonstrated the efficiency of DL compared to MLB approaches for AMC, less attention has been given to the UQ capabilities of these methods, which is the focus of our work.

More recently, Bayesian neural networks (BNNs) \cite{ref2} have been proposed to perform DL-based AMC with UQ capabilities. Although BNNs seemingly provide the advantage of both the high classification performance of DL as well as UQ capabilities, their practical application is hindered by lower classification performance compared to state-of-the-art CNNs. Despite their ability to capture predictive distribution and variance, which aids in uncertainty estimation, BNNs often have significantly lower classification performance compared to CNNs for AMC. This gap in performance and adaptability motivates our development of ensemble-based UQ methods for AMC. 

Ensemble learning has emerged as a compelling alternative to BNNs in multiple domains, such as image processing \cite{ref12} and bioinformatics \cite{de_bi}, where multiple models are jointly analyzed. Compared to single models, deep ensembles yield superior classification performance \cite{ref45} and generate predictive distributions instead of point estimates, leading to well-calibrated UQ estimates \cite{ref12, ref46, ref18}. Additionally, deep ensembles have demonstrated higher classification accuracy in the presence of data distortions such as OOD samples, which have been shown to limit the efficacy of DL-based AMC methods \cite{amc_adv_atk, amc_adv_atk2, ref40}.

Despite these successes in various other domains, the use of ensemble learning for UQ in AMC, or for signal processing tasks in general, remains unexplored. In this work, we leverage ensemble-based methods to demonstrate their effectiveness in quantifying uncertainty in AMC. Our framework not only achieves robust performance across a range of modulation types but is also scalable to state-of-the-art DL models, highlighting its adaptability in existing DL-based AMC frameworks.

\vspace{-0.5cm}

\section{Methodology} \label{methods}

In this section, we develop our deep ensemble-based AMC framework and describe the metrics used to quantify the uncertainty of each model. We begin by discussing our AMC signal model in Sec. \ref{sec:signal_model}. We then present our deep ensemble approach in Sec. \ref{sec:ensemble_modeling} and its complexity analysis in Sec. \ref{sec:cpx}. Subsequently, Sec. \ref{sec:estimation_metrics} details the estimation metrics employed to evaluate the performance of different methods. Finally, Sec. \ref{sec:ood_samples} discusses our approach for analyzing OOD AMC signals. 

\vspace{-0.25cm}

\subsection{Signal Modeling}
\label{sec:signal_model}

In our considered wireless communication environment, a transmitter transmits $\mathbf{s} =[s[0], \cdots, s[\ell - 1]]$ through a channel $\mathbf{h} \in \mathbb{C}^{\ell}$, where $\mathbf{h} =[h[0], \cdots, h[\ell - 1]]^{T}$ captures radio imperfections and selective fading and $\ell$ denotes the length of the received signal's observation window. We model the received signal as
\begin{equation}
    \mathbf{r} = \sqrt{\rho}\mathbf{H}\mathbf{s} + \mathbf{n},
\end{equation}
where $\mathbf{H} = \text{diag}\{h[0], \cdots, h[\ell - 1]\} \in \mathbb{C}^{\ell \times \ell}$, $\mathbf{n} \in \mathbb{C}^{\ell}$ represents additive white Gaussian noise (AWGN), and $\rho$ denotes the signal to noise ratio (SNR) of the received signal. Furthermore, we map each received baseband signal to a two-dimensional real matrix, $\mathbf{r} \in \mathbb{C}^{\ell} \rightarrow \mathbf{r} \in \mathbb{R}^{\ell \times 2}$ where the first and second column of $\mathbf{r}$ represent the in-phase and quadrature components, respectively, of $\mathbf{r}$ for compatibility with real-valued neural networks. 

The receiver aims to perform AMC by calculating $\text{argmax}_{i}P(m_{i} | \mathbf{r}, \theta)$, where $\theta$ parameterizes the model used to calculate the AMC probability (further discussed in Sec. \ref{sec:ensemble_modeling}), $m_{i} \in \mathcal{M}$, and $\mathcal{M} = \{m_{1}, \cdots, m_{C}\}$ represents the set of considered modulation constellations. We assume that the receiver uses $\mathcal{X}_{\text{tr}} = \{\mathbf{r}(n), \mathbf{y}(n); n = 1, \cdots,N\}$ and $\mathcal{X}_{\text{te}} = \{\mathbf{r}(t), \mathbf{y}(t); t = 1, \cdots,T\}$ as the training and testing datasets, respectively, where $\mathcal{X}_{\text{tr}} \cap \mathcal{X}_{\text{te}} = \emptyset$.

\subsection{Ensemble Modeling}
\label{sec:ensemble_modeling}

Data-driven AMC models, alone, cannot effectively characterize the uncertainty associated with their predictions because they tend to produce overconfident outputs. To address this limitation, we adopt an ensemble modeling approach consisting of multiple state-of-the-art deep learning AMC classifiers \cite{ref52}. Unlike a standalone model, ensemble predictions cannot be obtained directly because multiple models are used simultaneously. To ensure diversity in predictions, each classifier is initialized with random parameters while maintaining the same architecture, resulting in each model converging to different local minima thus ensuring parallelized training as well as diverse predictions despite training on the same dataset \cite{ref11}. As a result, the ensemble exhibits variability, yielding varying levels of confidence on samples that are known to be difficult for deep learning classifiers to operate on such as low SNR signals. As a result, analyzing the joint output prediction from each classifier in the ensemble simultaneously improves the UQ of the model. 

We denote our ensemble as $\theta = \{\theta_b\}_{b=1}^{B}$ where $B$ denotes the number of classifiers in the ensemble and $\theta_{b}$ parameterizes the $b^{\text{th}}$ model. 
Each AMC classifier in the ensemble is a deep learning model with a softmax output. We denote each deep learning classifier as $f_{\theta_{b}}(\textbf{r}): \mathbb{R}^{\ell \times 2} \rightarrow \mathbb{R}^{C}$, parameterized by $\theta_{b}$. Here, the $b^{\text{th}}$ model aims to map the received signal $\mathbf{r} \in \mathbb{R}^{\ell \times 2}$ to a modulation constellation $\mathbf{y} \in \mathbb{R}^{C}$, where $C = |\mathcal{M}|$ represents the total number of possible modulation constellations. Due to the softmax classifier, $f_{\theta_{b}}(\textbf{r})$ outputs $\hat{\mathbf{y}}^{(b)} \in \mathbb{R}^{C}$, where $\hat{\mathbf{y}}_{i}^{(b)} = P(m_{i} | \mathbf{r}, \theta^{(b)}) \in \mathbb{R}$ (i.e., $\hat{\mathbf{y}}_{i}^{(b)}$ is the $i^{\text{th}}$ element of $\hat{\mathbf{y}}^{(b)}$ and denotes the probability assigned by the $b^{\text{th}}$ classifier that the input $\mathbf{r}$ is modulated according to constellation $m_{i}$). 
We train each AMC classifier in the ensemble by optimizing
\begin{equation}
\underset{\theta_{b}}{\text{min}}\hspace{1mm} \mathcal{L}(\theta_b,\textbf{r}_,\mathbf{y}), 
\end{equation}
where 
\begin{equation}
    \mathcal{L}(\theta_b,\textbf{r}_,\mathbf{y}) =  -\frac{1}{N} \sum_{i=1}^{N} \sum_{j=1}^{C} \mathbf{y}_{ij} \log(\hat{\mathbf{y}}_{ij}), \label{cce_loss}
\end{equation}
$\mathbf{y}_{ij} \in \mathbb{R}$ is $j^{\text{th}}$ element of $\mathbf{y} \in \mathbb{R}^{C}$ corresponding to the $i^{\text{th}}$ signal, and $\hat{\mathbf{y}}_{ij} \in \mathbb{R}$ is $j^{\text{th}}$ element of $\hat{\mathbf{y}} \in \mathbb{R}^{C}$, the predicted label, corresponding to the $i^{\text{th}}$ signal. 

During evaluation, we jointly analyze the prediction from all $B$ classifiers in the ensemble to obtain 
\begin{equation} \label{agg}
    P(m_{i} | \mathbf{r}, \theta) = \frac{1}{B} \sum_{b=1}^{B}P(m_{i} | \mathbf{r}, \theta^{(b)}), 
\end{equation}
where $\text{argmax}_{i}P(m_{i} | \mathbf{r}, \theta)$ yields the predicted modulation constellation from the ensemble. In addition to obtaining the probability with which the input belongs to a particular modulation classification, we use the distributive estimate produced by the ensemble to characterize additional UQ metrics, which are discussed in Sec. \ref{sec:estimation_metrics}.

\subsection{Complexity Analysis}
\label{sec:cpx}

Here, we analyze the computational complexity of our proposed framework during inference. The computational complexity of a single layer, $l$, of a CNN during runtime is given by $\mathcal{O}(C_{\text{in}}^{(l)} \cdot C_{\text{out}}^{(l)} \cdot K_{w}^{(l)} \cdot K_{h}^{(l)} \cdot H^{(l)} \cdot W^{(l)})$ and, thus, the complexity over an entire CNN is given by the sum of these complexities, resulting in
\begin{equation} \label{comp_cpx}
\sum_{l=1}^{L} \mathcal{O} \bigg{(}C_{\text{in}}^{(l)} \cdot C_{\text{out}}^{(l)} \cdot K_{w}^{(l)} \cdot K_{h}^{(l)} \cdot H^{(l)} \cdot W^{(l)}\bigg{)},
\end{equation}
where $L$ is the number of layers and, in each layer $l$, $C_{\text{in}}^{(l)}$ is the number of input channels, $C_{\text{out}}^{(l)}$ is the number of output channels, $K_{w}^{(l)}$ and $K_{h}^{(l)}$ are the width and height of each convolutional kernel, and $H^{(l)}$ and $W^{(l)}$ represent the output spatial size. Since each CNN in our proposed framework is independent, they can process each received signal in parallel and thus operate without incurring any additional computational overhead compared to a single CNN. Thus, the computational complexity of our proposed method is given by (\ref{comp_cpx}). Moreover, in terms of wall clock time, each sample requires, on average, 1 ms per sample to process. Thus, our proposed framework does not incur any additional overhead in comparison to state-of-the-art deep learning-based AMC approaches.

\subsection{Estimation Metrics}
\label{sec:estimation_metrics}

Beyond obtaining the modulation classification estimate given in (\ref{agg}), we also employ several proper UQ scoring metrics \cite{ref12} to assess model performance comprehensively. In this capacity, we consider the negative log-likelihood (NLL), the Brier score, the width of the prediction's confidence interval (CI), the coverage of the prediction, the set of high-confidence predictions, the expected calibration error (ECE), and the kullback-leibler (KL) divergence.

The NLL quantifies the likelihood of the predicted probabilities aligning with the true labels and is given over the entire testing set by
\begin{equation}
    -\frac{1}{T} \sum_{t=1}^{T} \sum_{j=1}^{C} \mathbf{y}_{tj} \log(\hat{\mathbf{y}}_{tj}),
\end{equation}
where lower NLL values indicate that the model assigns higher confidence to the correct classes. In AMC, low NLL is important because poor model training under low SNR conditions can lead to overconfident misclassifications. Thus, NLL serves as an indicator of probabilistic robustness under channel impairments.

Similarly, the Brier score, which measures the mean squared difference between the predicted probabilities and the actual outcomes, is computed over all samples and classes and is given by
\begin{equation}
    \frac{1}{T} \sum_{t=1}^{T} \sum_{j=1}^{C} (\mathbf{y}_{tj} - \hat{\mathbf{y}}_{tj})^2, 
\end{equation}
where lower Brier score values indicate that the model's probabilistic predictions are closer to ground truth values, minimizing deviations. Unlike accuracy, the Brier score penalizes both overconfidence in incorrect predictions and under confidence in correct predictions. In AMC systems, where prediction certainty impacts communication reliability, the Brier score provides an interpretable measure of prediction quality.

In addition, we also analyze the Confidence Interval (CI) widths, which characterize the ensemble's uncertainty in each prediction. A higher CI width indicates higher uncertainty, as the models comprising the ensemble vary to a higher degree in their predictions, whereas lower CI widths indicate lower uncertainty as the majority of models agree about the predicted modulation class.  Ideally, we aim for a wide range of CI widths rather than a concentration within a narrow range, as seen with singular deep learning models, which indicates overconfidence. The upper and lower bounds of the CI of the $t^{\text{th}}$ sample belonging to the $j^{\text{th}}$ constellation are given by 
\begin{equation}
    \hat{\mathbf{y}}_{tj} \ \pm\ z_{\alpha} \cdot \sqrt{\frac{S^2}{B}}, \label{ci_ul}
\end{equation}
$z_{\alpha}$ is the $1 - \alpha / 2$ quantile of a zero-mean unit variance Gaussian distribution, and $S^{2}$ is the variance of the predicted probability among models, and the CI width is given by computing the difference between the upper and lower bounds of (\ref{ci_ul}). In AMC applications, CI widths offer a direct measure of the ensemble's epistemic uncertainty, resulting in a quantified interpretable measure of the confidence associated with each signal's modulation constellation prediction.

We next calculate the coverage proportion, which evaluates the number of samples in the testing set whose true label is within the $z_{\alpha}$ CI. In computing the coverage of the ensemble, we consider two scenarios. First, we employ a strict condition, where the CI of the true class must contain one and the CI of all other classes must contain zero. This scenario strictly requires the ensemble to be certain that the predicted sample simultaneously (i) is modulated according to the predicted class and (ii) is not modulated according to any other class. Second, we relax the second condition and only require the CI of the true class to contain one and we do not consider the CIs of the other classes. This metric, in both scenarios, quantifies the proportion of samples for which the ensemble is confident at a $z_{\alpha}$ level that the input belongs to the predicted class. The coverage is especially helpful for understanding the ensemble's confidence under low SNR conditions, which often is the cause of lowered classification performance in AMC applications. It is important to note that such confidence intervals can extend beyond the $[0,1]$ range particularly when the mean is near zero or one, indicating that the model is generally confident in its prediction about the correct class while simultaneously being confident that the sample does not belong to another class (i.e., the strict condition) or merely being confident in its prediction about the correct class without considering the confidence of other incorrect classes (i.e, the relaxed condition).

To explore confidence further, we define the high confidence set to evaluate the frequency with which the ensemble is highly confident in its predictions. A high confidence prediction occurs when $\text{argmax}_{i}P(m_{i} | \mathbf{r}, \theta) > 0.8$. A high confidence prediction indicates overconfidence, which often occurs in singular deep learning models, while a low confidence prediction reflects under-confidence. Ideally, in AMC applications, the ensemble should strike a balance, avoiding both extremes. This metric, like others previously discussed, provides a measure for model reliability during adverse channel conditions in AMC applications.

We also compute the Expected Calibration Error (ECE), which  measures how far the model’s predicted probabilities are from the true frequency of correct predictions. The ECE is given by
\begin{equation}
\sum_{k=1}^{\mathcal{K}}
\frac{\lvert \zeta_{k}\rvert}{T}
\;\bigl|\mathrm{acc}(\zeta_{k}) - \mathrm{conf}(\zeta_{k})\bigr|,
\end{equation}
where, for each bin $\zeta_{k}$ we take the fraction of samples in that bin, $\lvert \zeta_{k}\rvert / T$, multiply it by the absolute difference between the bin’s actual accuracy $\mathrm{acc}(\zeta_{k})$ and its average predicted confidence $\mathrm{conf}(\zeta_{k})$, and then sum these values over all $\mathcal{K}$ bins to yield the ECE. A lower ECE indicates that the model’s prediction estimates are well‐calibrated to its actual accuracy.

Finally, we measure the Kullback–Leibler (KL) divergence between the one-hot true distribution $\mathbf{y}_{t}$ and the predictive distribution $\hat{\mathbf{y}}_{t}$ for each sample, the KL divergence is given by
\begin{equation}
D_{\mathrm{KL}}\bigl(\mathbf{y}_{t} \,\big\|\, \hat{\mathbf{y}}_{t}\bigr)
\;=\;
\sum_{j=1}^{C} y_{tj}\,\log\!\Bigl(\tfrac{y_{tj}}{\hat{y}_{tj}}\Bigr).
\end{equation}
Here, lower KL divergence indicates that the model assigns higher probability to the true modulation class, yielding more accurate distributions. In AMC, this reflects both the confidence and correctness of probabilistic predictions under noisy channel conditions.

\subsection{Out-of-distribution Samples}
\label{sec:ood_samples}

To further assess the robustness of our framework and simulate real-world environmental factors, we generate adversarial examples, denoted as $\textbf{r}'$, which are considered OOD samples \cite{ref56}. Deep learning AMC classifiers are highly susceptible to adversarial examples as they induce erroneous predictions, reflected in the softmax output of standalone models, particularly on high SNR signals. Although approaches to mitigate such samples have been investigated \cite{ade,amc_adv_trn,amc_def2}, they specifically target improving the robustness to adversarial attacks and do not consider the general UQ capabilities of the model. These samples degrade classification performance by pushing the received signal across the decision boundaries of DL models. This phenomenon is often represented as an adversarial perturbation, $\boldsymbol{\delta} \in \mathbb{R}^{\ell \times 2}$, which is introduced to existing samples and formulated as
\begin{equation}
    \min_{\boldsymbol{\delta}} \|\boldsymbol{\delta}\|_{\infty} \nonumber
\end{equation}
\begin{equation}
    \text{s.t.} \quad f_{\theta_{b}}(\mathbf{r}) \neq f_{\theta_{b}}(\mathbf{r} + \boldsymbol{\delta}) \nonumber
\end{equation}
\begin{equation}
    \mathbf{r} + \boldsymbol{\delta} \in \mathbb{R}^{\ell \times 2} \label{adv_obj}
\end{equation}
where $\textbf{r}' = \textbf{r} + \boldsymbol{\delta}$ represents the perturbed version of $\mathbf{r}$ and $\|\cdot\|_{\infty}$ denotes the $\ell_{\infty}$ norm bound of $\boldsymbol{\delta}$. Note that adversarial attacks using other norm bounds (e.g., the $l_{0}$, $l_{1}$, or $l_{2}$ bound) can also be generated, but we select the $l_{\infty}$ bound as it perturbs each sample thus helping in shift the sample away from its true distribution, which is our overall objective. 

In practice it is difficult to analytically solve (\ref{adv_obj}) due to its excessive non-linearity. Thus, solution to (\ref{adv_obj}) are approximated in practice. In this work, we utilize the Fast Gradient Sign Method (FGSM) \cite{ref51}, to approximate a solution to (\ref{adv_obj}), which is given by
\begin{equation}
    \mathbf{r}' = \mathbf{r} + \epsilon \, \text{sign}(\nabla_{\mathbf{r}} \mathcal{L}(\theta_{b}, \mathbf{r}, \mathbf{y})),
\end{equation}
where $\epsilon = ||\boldsymbol{\delta}||_{\infty}$ represents the $l_{\infty}$-bounded perturbation magnitude applied to the input samples.

We quantify the effect of the additive perturbation using the perturbation-to-noise ratio (PNR) \cite{amc_adv_atk}. Here, we compute the expected value of the power of the perturbation ${\mathbb{E}\left[\|\boldsymbol{\delta}\|_2^2\right]}$ and then obtain the expected value of the power of the received signal ${\mathbb{E}\left[\|\textbf{r}\|_2^2\right]}$. Using these quantities, along with the SNR of the received signal, the PNR is given by 
\begin{equation}
    \text{PNR [dB]} = \frac{\mathbb{E}\left[\|\boldsymbol{\delta}\|_2^2\right]}{\mathbb{E}\left[\|\textbf{r}\|_2^2\right]} \, \text{[dB]} + \text{SNR [dB]}, 
\end{equation}
where a lower PNR indicates that the perturbation is less potent and has a smaller effect on the distribution shift of $\mathbf{r}$ whereas a higher PNR shift $\mathbf{r}'$ further from $\mathbf{r}$ and is more likely to satisfy the constraints of (\ref{adv_obj}). 

\section{Performance Evaluation}
\label{sec:performance_evaluation}

In this section, we assess the effectiveness of our proposed UQ framework for AMC. We begin by detailing our experimental setup in Sec. \ref{sec:exp_setup}. In Sec. \ref{sec:uq_capabilities}, we provide an in-depth analysis of the uncertainty estimation capabilities of our proposed framework compared to the proposed baselines. In Sec. \ref{sec:ood_performance}, we investigate our framework’s robustness to OOD samples and adversarial perturbations, a critical consideration for deploying AMC systems in real-world, dynamic environments. Finally, in Sec. \ref{sec:complexity analysis} we provide a discussion as well as insights highlighting the benefits of our proposed method.


\begin{figure}[t]
    \centering
    \includegraphics[width=0.98\linewidth,height=0.98\linewidth]{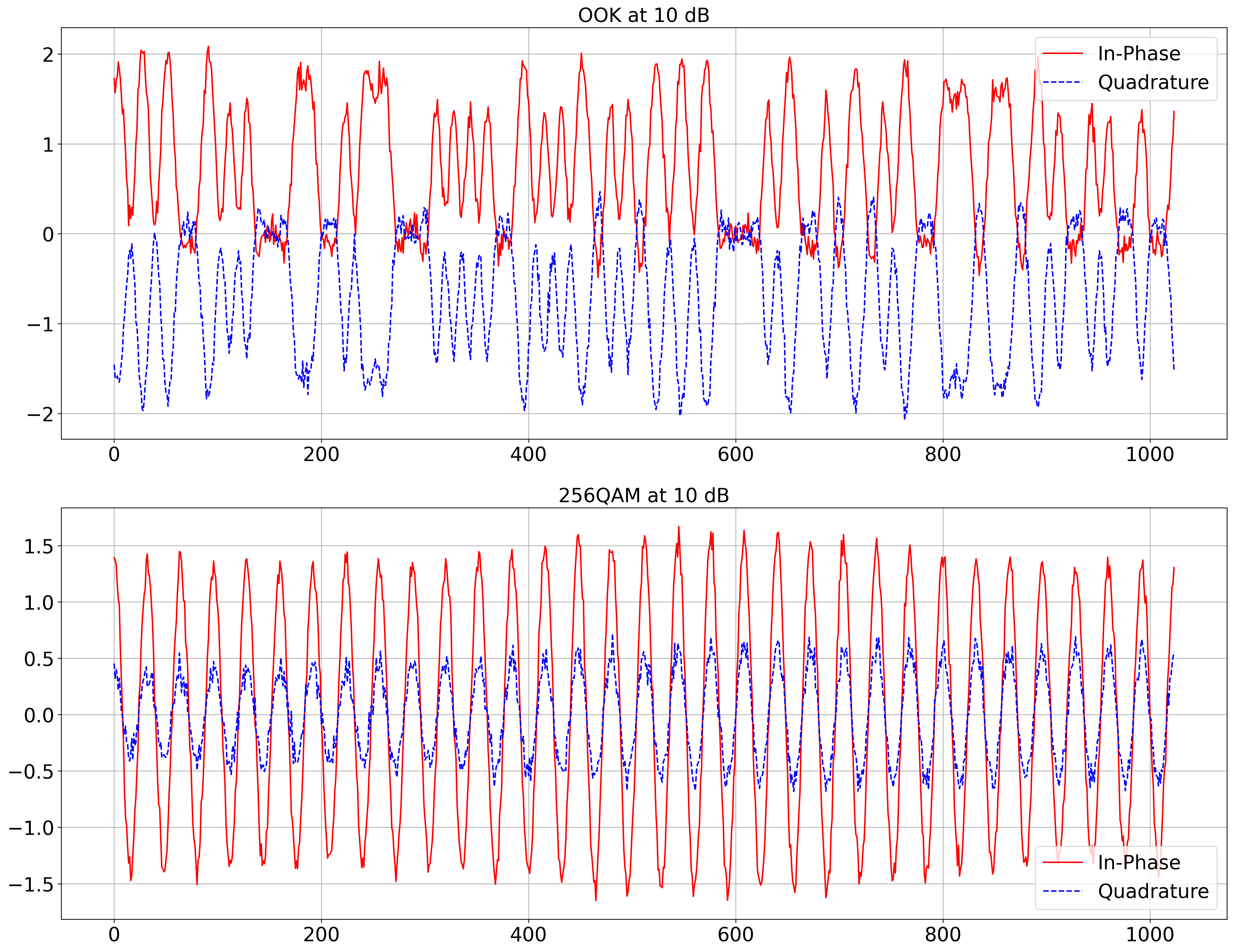}
    \caption{Sample modulation constellations used for AMC in our empirical evaluations including OOK (top) and 256QAM (bottom) from the RML 2018.01a dataset.}
    \label{figSignal}
\end{figure}

\subsection{Experimental Setup}
\label{sec:exp_setup}

We perform our empirical evaluation on two AMC datasets: RadioML2016.10a \cite{ref31} and RadioML2018.01a \cite{ref32}. RadioML2016.10a is a synthetic dataset generated using GNU radio while RadioML2018.01a is collected over-the-air on a wireless testbed making it a more difficult dataset but a better representation of real-world AMC data. RadioML2016.10a is comprised of $C = 11$ modulation types ranging from $-10$ dB to $20$ dB SNR in increments of $2$ dB. For each SNR, the dataset contains $12,000$ signals, where $\ell = 128$ samples in length. In comparison, RadioML2018.01a is an over-the-air dataset that contains $C = 24$ modulation types, ranging from $-20$ dB to $18$ dB in increments of $2$ dB, with each signal consisting of $\ell = 1024$ samples and consists of $19,661$ signals per SNR.  Fig. \ref{figSignal} visualizes selected constellations used in our analysis. In each experiment, we use an 80\%/20\% split of the signals at each SNR to form $\mathcal{X}_{\text{tr}}$ and $\mathcal{X}_{\text{te}}$. $\mathcal{X}_{\text{tr}}$ is used to train each classifier in the ensemble. After training, we report the UQ characteristics of the ensemble on $\mathcal{X}_{\text{te}}$. 

Our framework employs an ensemble of $B = 15$ CNNs, selected for their demonstrated effectiveness in AMC tasks \cite{ref9}. All models share the same architecture, which is shown in Table \ref{tab1} and are all independently trained on $\mathcal{X}_{\text{tr}}$ by minimizing (\ref{cce_loss}) using stochastic gradient descent (SGD). Each model in the ensemble is trained for 100 epochs with a batch size of 256 and a learning rate of 0.001. The hyperparameters were initially selected based on commonly used settings in the AMC literature \cite{ref_bs}, and they were further fine-tuned for our application using a grid search. Although we use the CNN architecture shown in Table \ref{tab1}, our framework can be extended to incorporate an ensemble of classifiers with any architecture. As we will show, our ensemble of AMC classifiers consistently outperforms a standalone AMC classifier with the same architecture. 

\begin{table}[t]
\centering
\caption{The CNN architecture of each classifier in our ensemble.}
\label{tab1}
\begin{tabular}  {| c | c | c | c |} 
\hline
\textbf{Layer} & \textbf{Dropout Rate (\%)} & \textbf{Activation} & \textbf{Shape} \\ 
\hline
Conv 1 & 20 & ReLU   & $3 \times 1 \times 256$ \\
Conv 2 & 20 & ReLU   & $3 \times 2 \times 128$ \\
Conv 3 & 20 & ReLU   & $3 \times 1 \times 64$ \\
Conv 4 & 20 & ReLU   & $3 \times 1 \times 64$ \\
Flatten   & -  & -   & - \\
Dense   & -  & ReLU   & 128 \\
Output & -  & Softmax & $24$ \\ 
\hline
\end{tabular}
\end{table}


\begin{figure}[t]
    \centering
    \subfloat{%
        \includegraphics[width=0.48\linewidth]{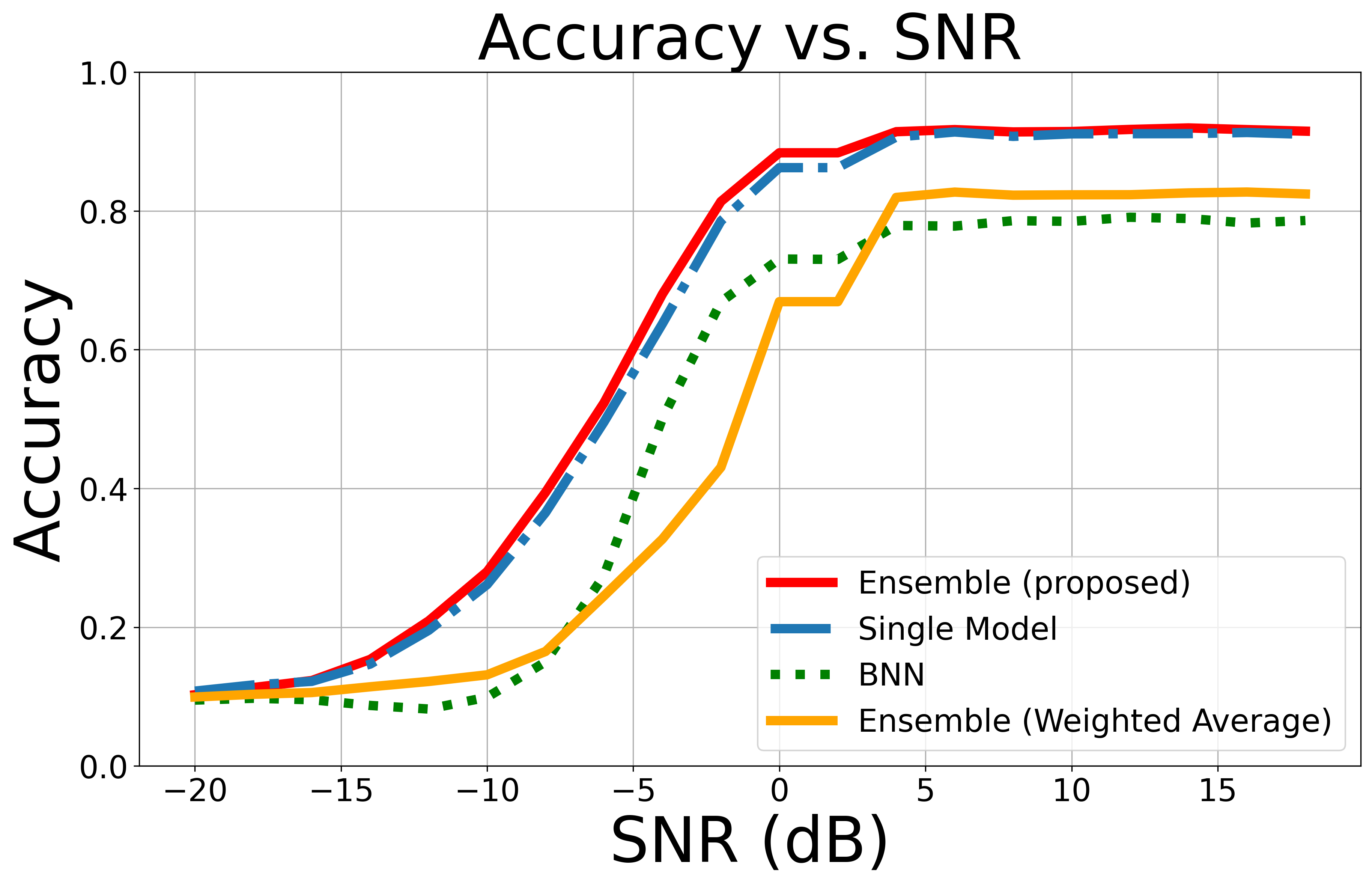}
        \label{fig:2016_acc}
    }
    \subfloat{%
        \includegraphics[width=0.48\linewidth]{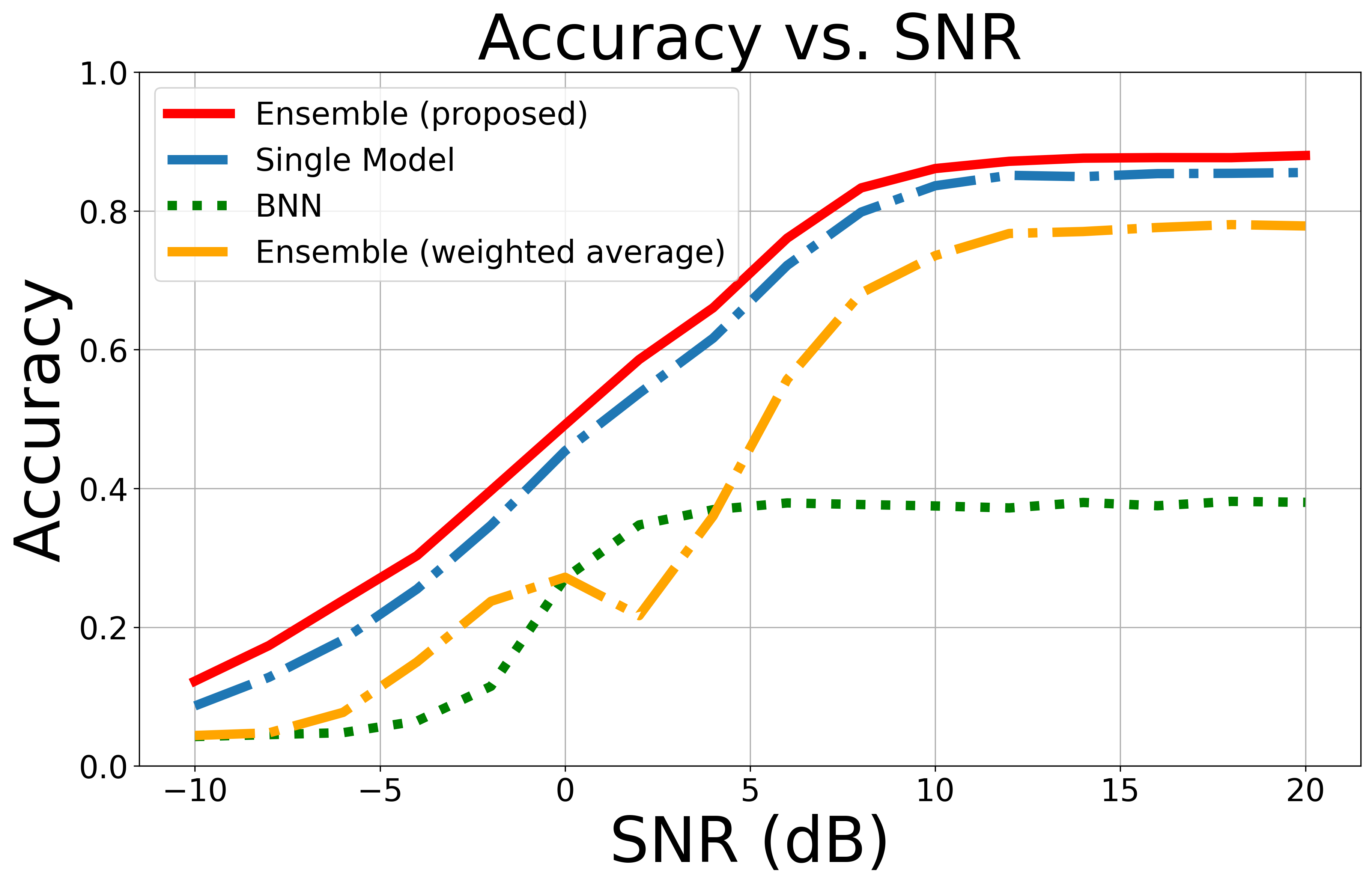}
        \label{fig:2018_acc}
    }
    
    \subfloat{%
        \includegraphics[width=0.48\linewidth]{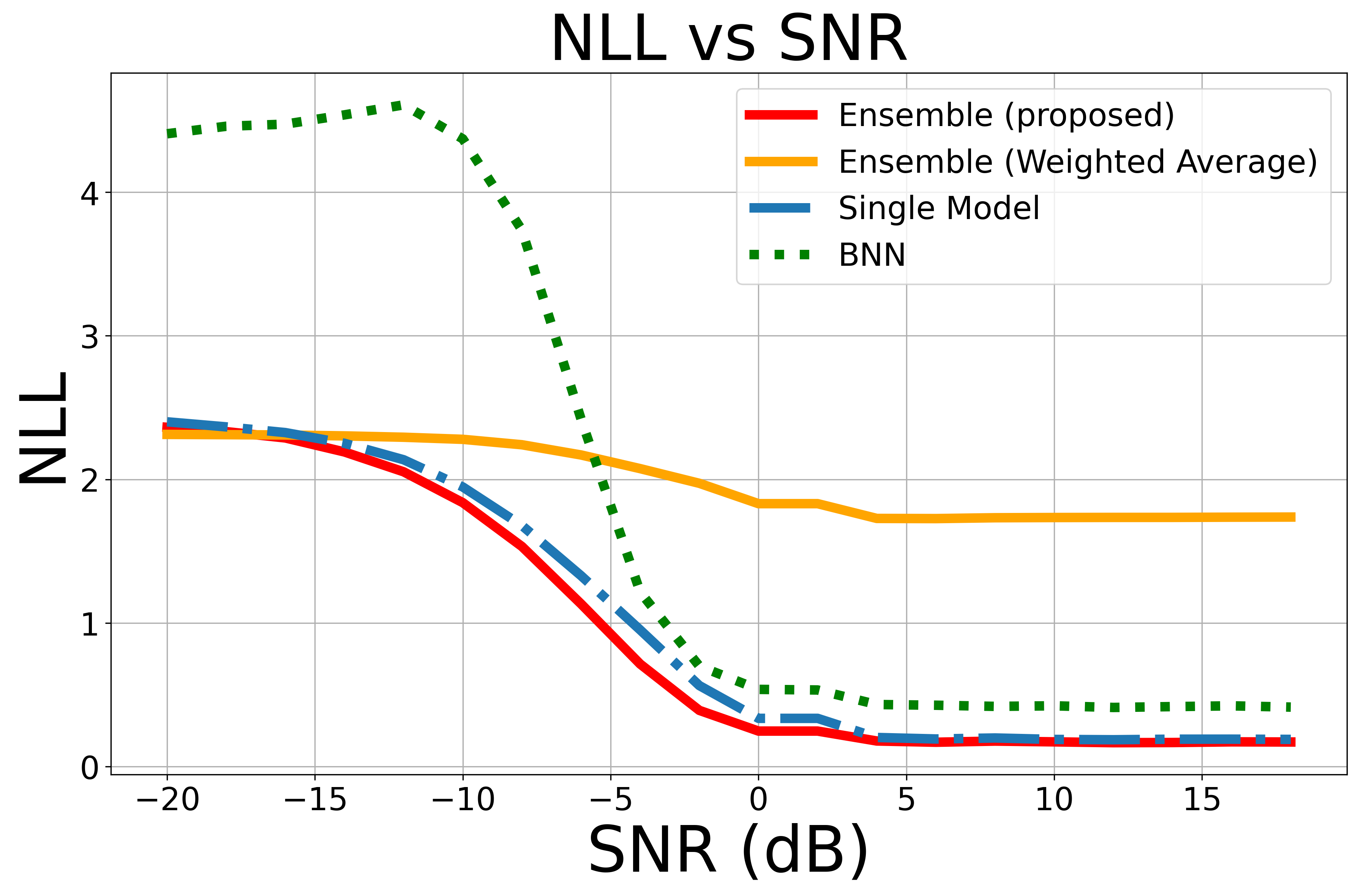}
        \label{fig:2016_nll}
    }
    \subfloat{%
        \includegraphics[width=0.48\linewidth]{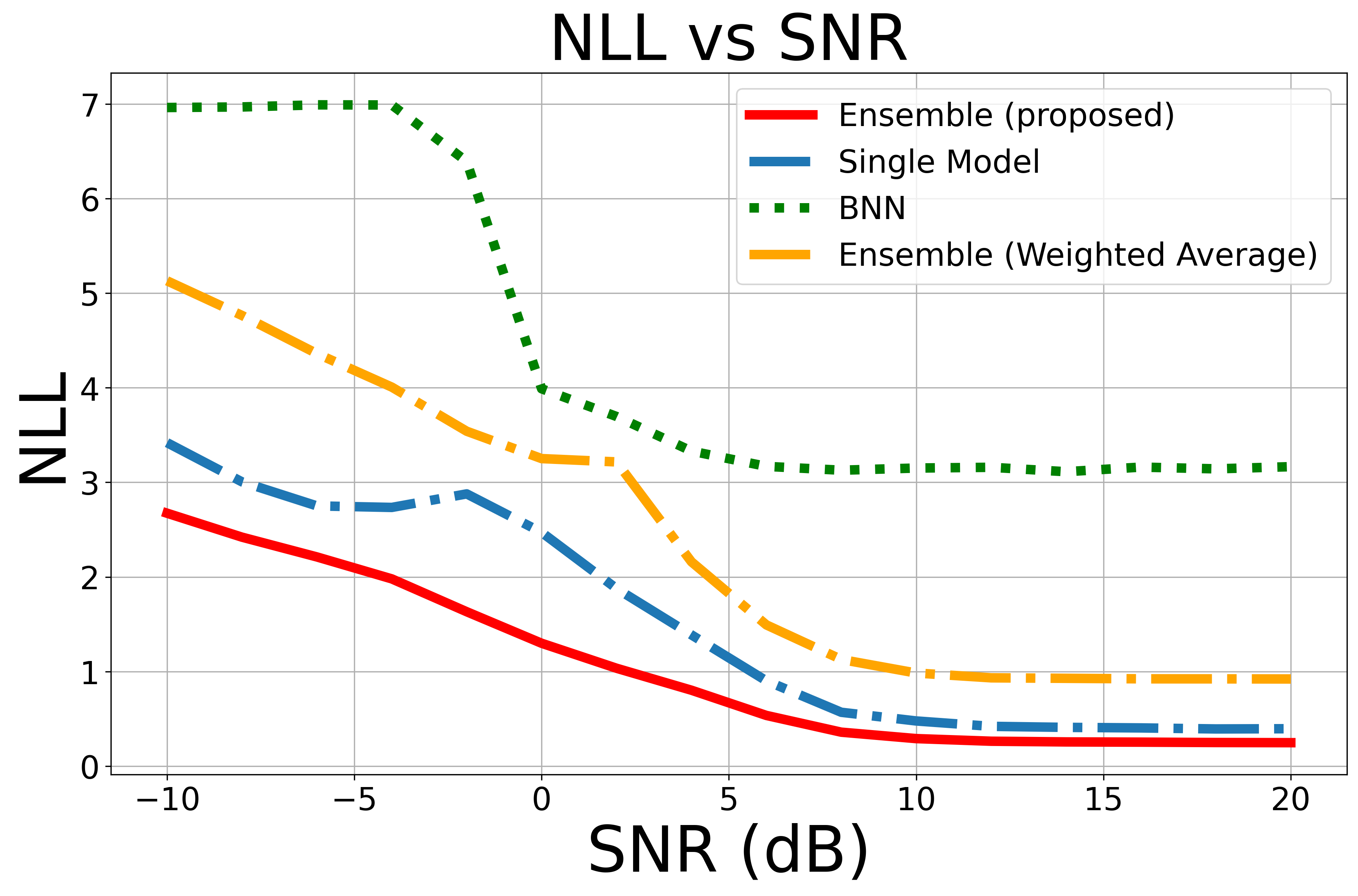}
        \label{fig:2018_nll}
    }
    
    \subfloat{%
        \includegraphics[width=0.48\linewidth]{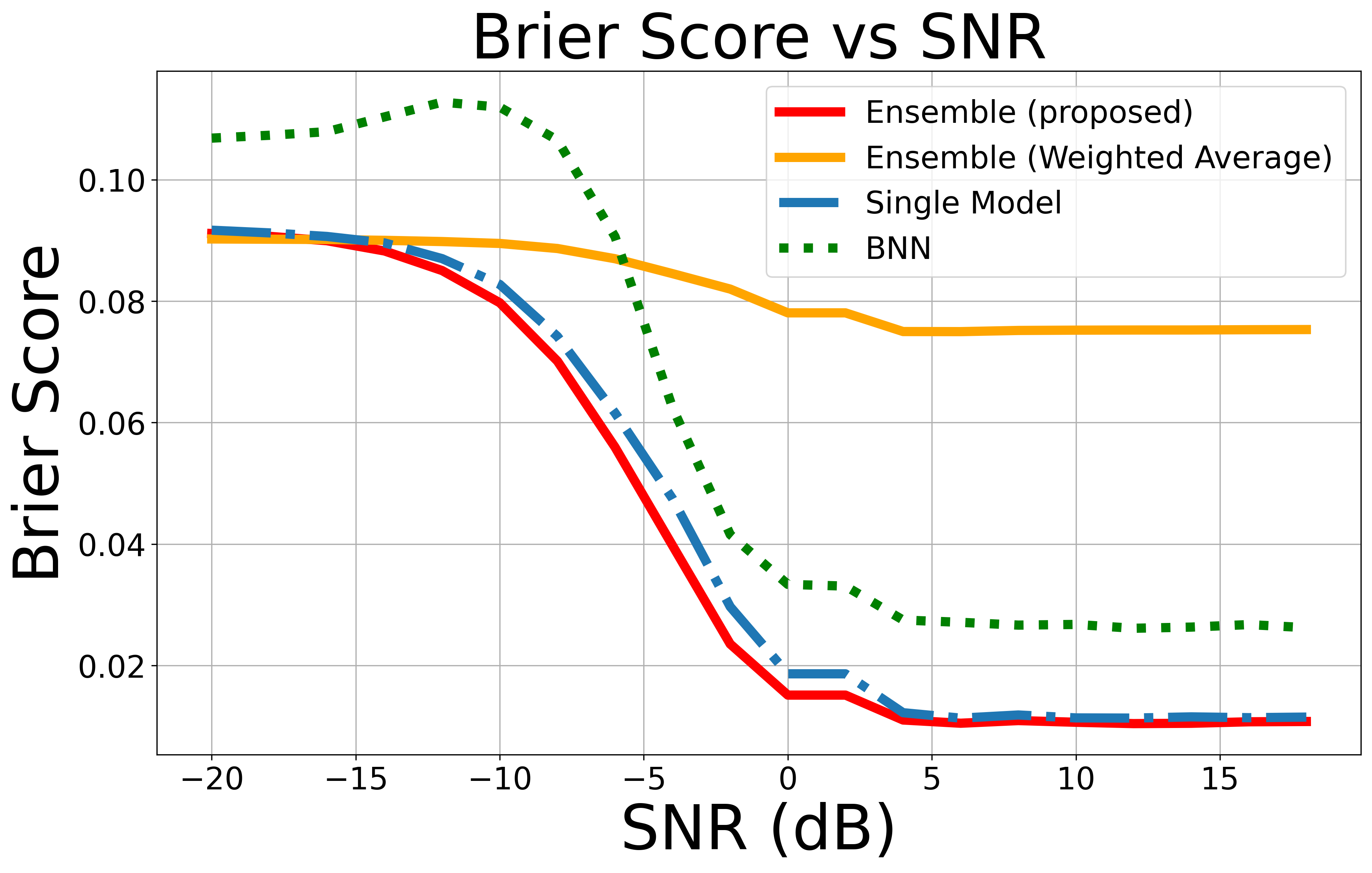}
        \label{fig:2016_brier}
    }
    \subfloat{%
        \includegraphics[width=0.48\linewidth]{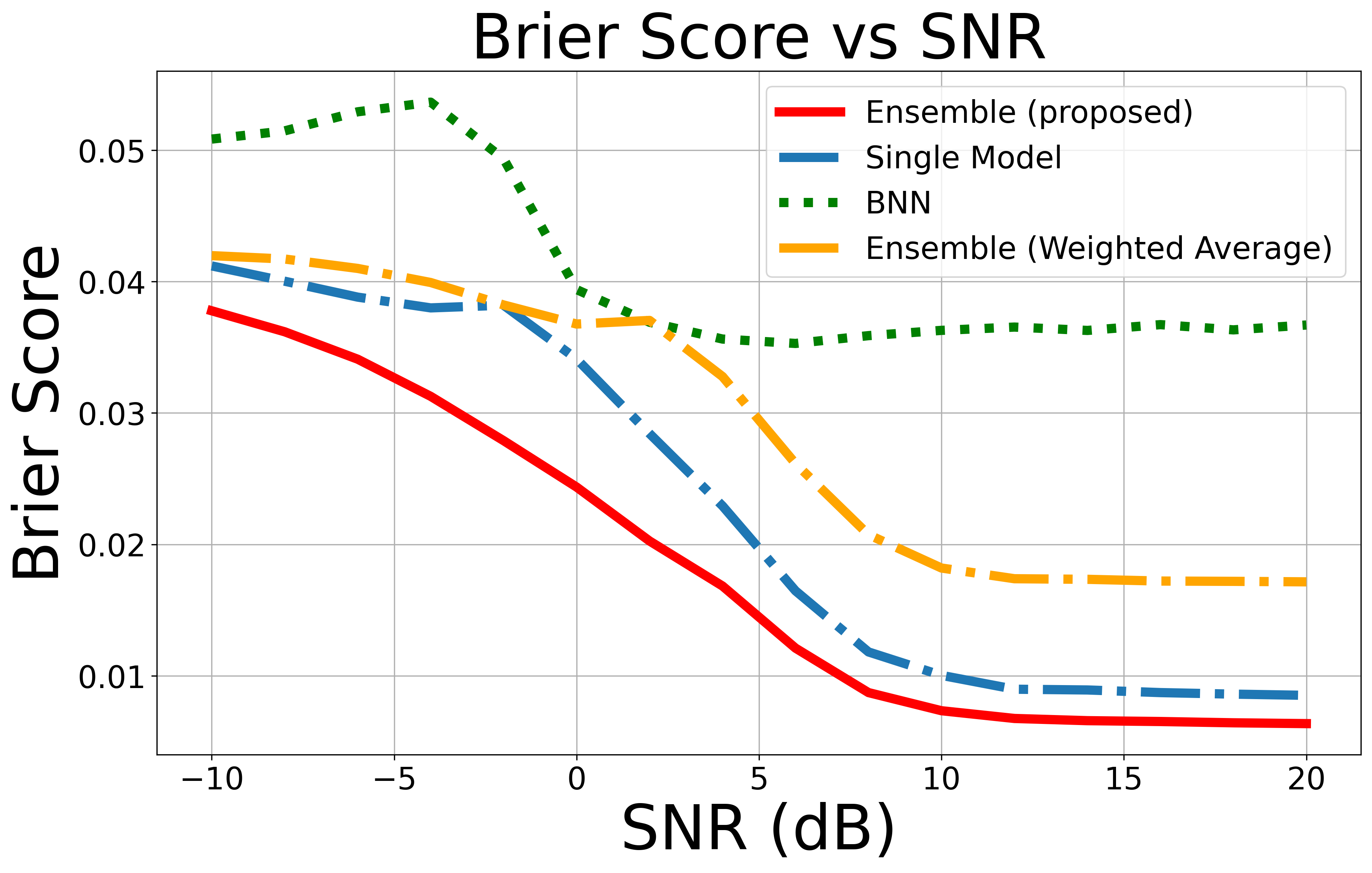}
        \label{fig:2018_brier}
    }

    \caption{UQ metrics on the 2016.10a (left column) and RML 2018.01a (right column) datasets. We see that our proposed ensemble model outperforms all baselines on each considered metric for both datasets.}
    \label{fig2}
\end{figure}



\begin{figure}[t]
    \centering
    \subfloat{%
        \includegraphics[width=0.33\linewidth, height=0.24\linewidth]{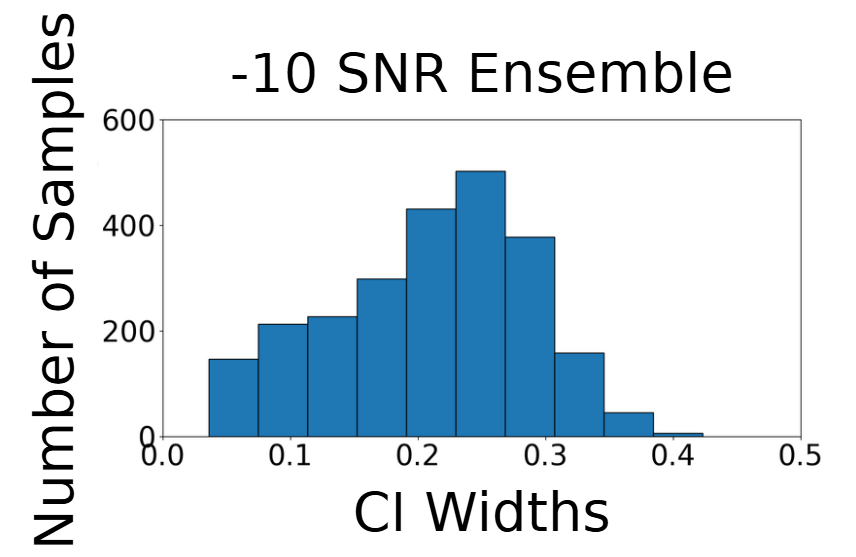}
        \label{fig:-10SNR_ensemble_C}
    }
    \subfloat{%
        \includegraphics[width=0.33\linewidth, height=0.24\linewidth]{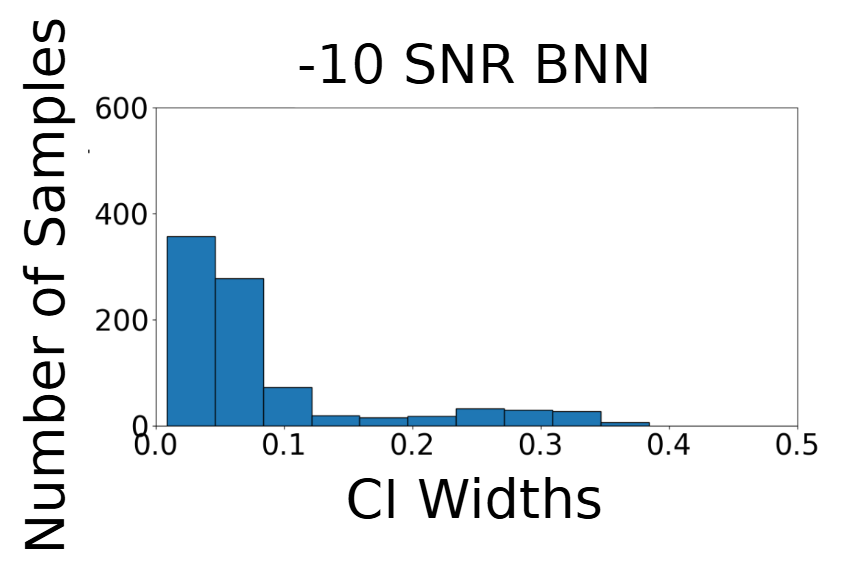}
        \label{fig:-10SNR_BNN_C}
    }
    \subfloat{%
        \includegraphics[width=0.33\linewidth, height=0.24\linewidth]{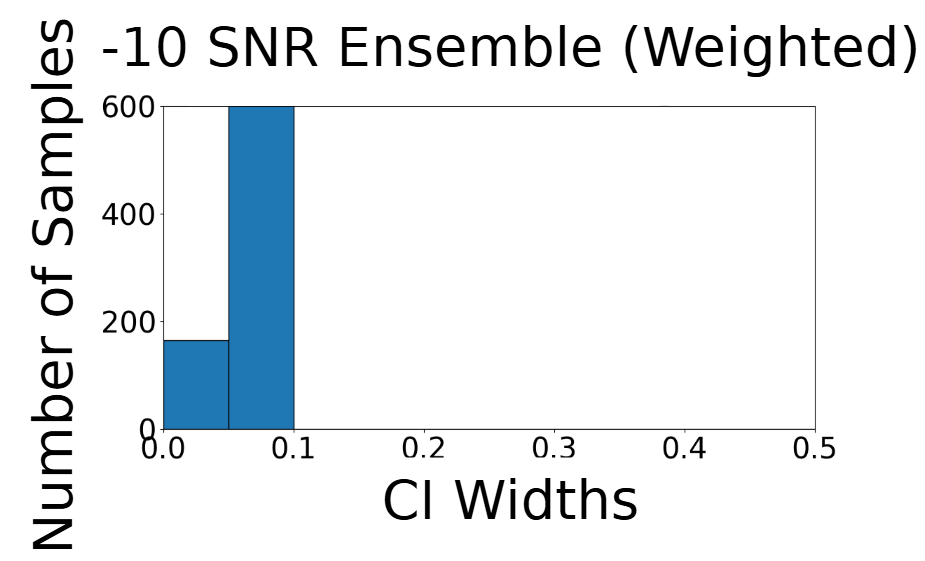}
        \label{fig:neg10SNR_WEN_C.png}
    }

    \subfloat{%
    \includegraphics[width=0.33\linewidth,height=0.24\linewidth]{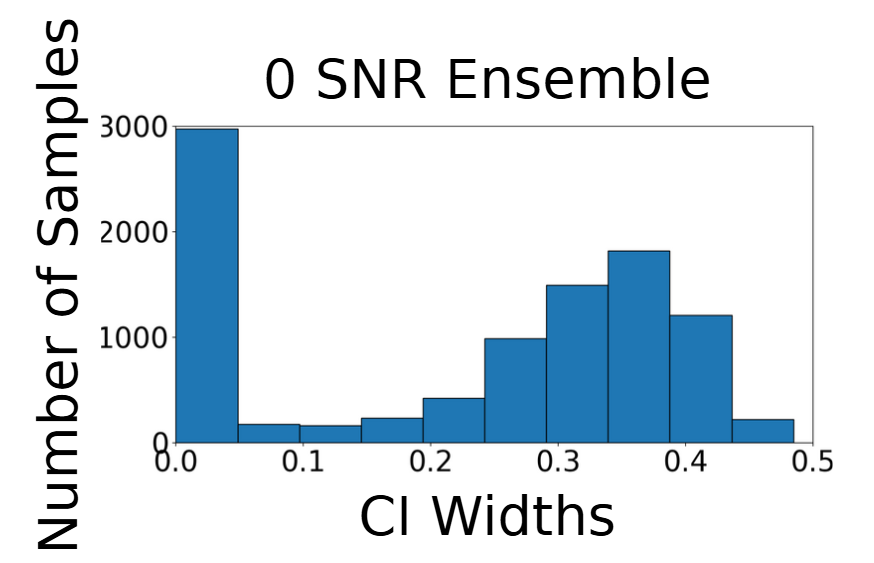}
        \label{fig:0SNR_ensemble_C}
    }
    \subfloat{%
        \includegraphics[width=0.33\linewidth,height=0.24\linewidth]{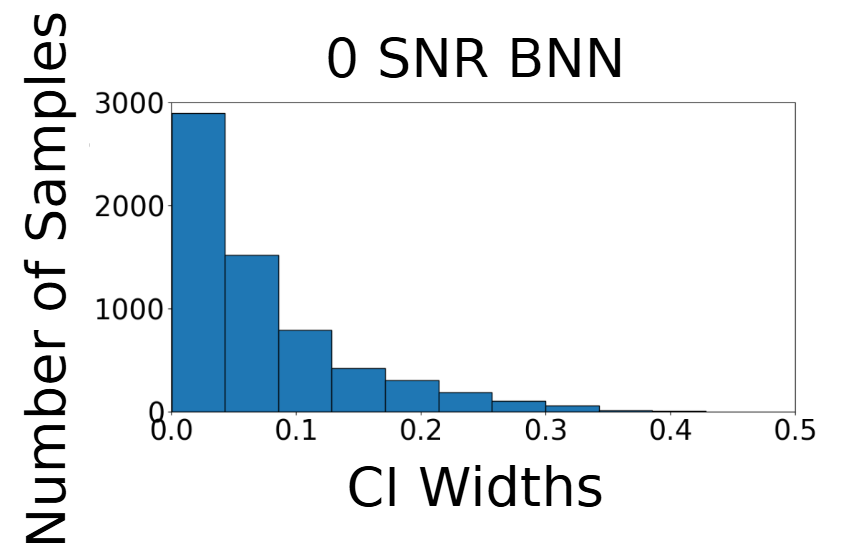}
        \label{fig:0SNR_BNN_C}
    }
    \subfloat{%
        \includegraphics[width=0.33\linewidth, height=0.24\linewidth]{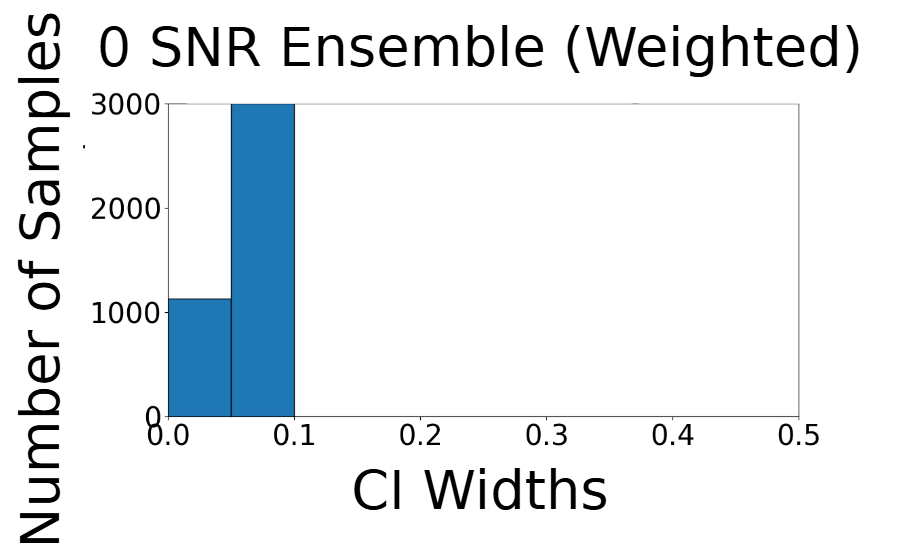}
        \label{fig:0SNR_WEN_C.png}
    }

    \subfloat{%
        \includegraphics[width=0.33\linewidth,height=0.24\linewidth]{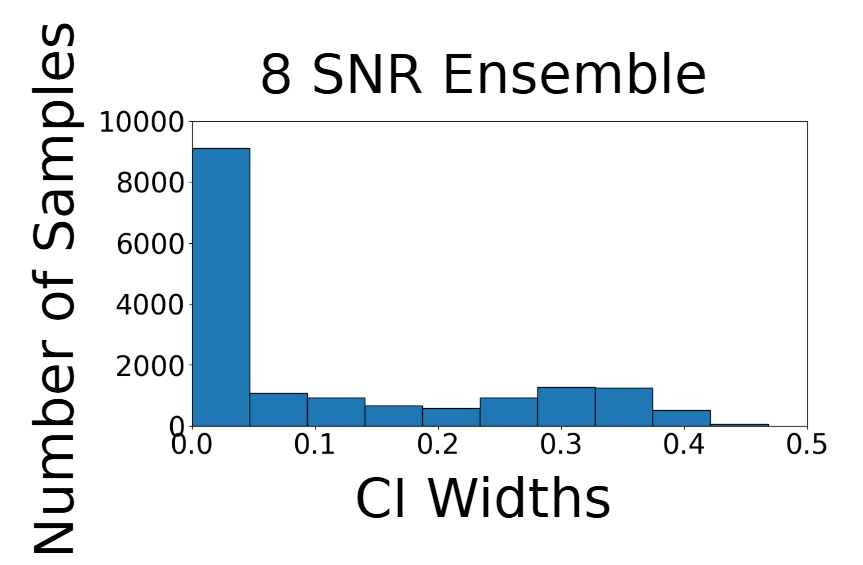}
        \label{fig:8SNR_ensemble_C}
    }
    \subfloat{%
        \includegraphics[width=0.33\linewidth,height=0.24\linewidth]{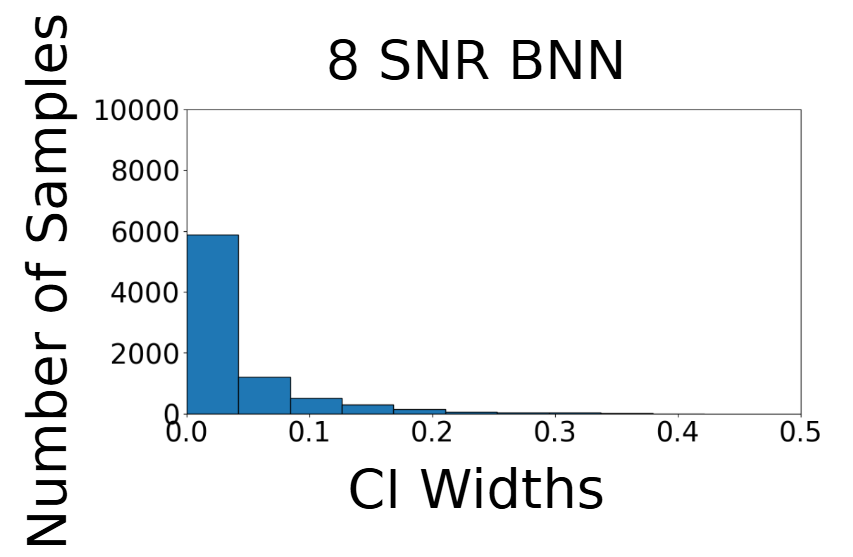}
        \label{fig:8SNR_BNN_C}
    }
    \subfloat{%
        \includegraphics[width=0.33\linewidth, height=0.24\linewidth]{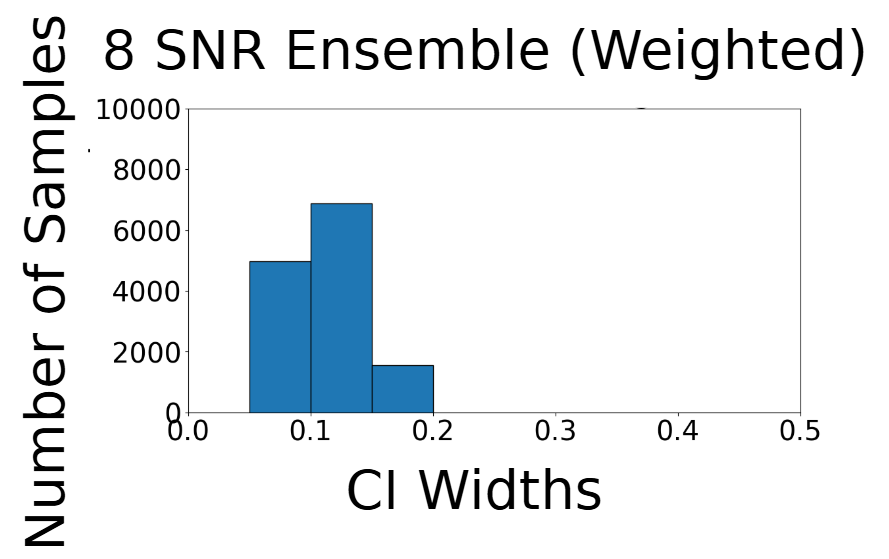}
        \label{fig:8SNR_WEN_C.png}
    }

    \caption{
    Correctly predicted CI widths of our proposed ensemble (left column), the baseline BNN approach (middle column), and the baseline weighted ensemble (right column) on RML 2018.01a. Here, we see that our ensemble has wider CI widths compared to both baselines, demonstrating that our proposed ensemble approach can characterize uncertainty to a higher extent in comparison to the BNN and weighted ensemble.}
    
    \label{fig:fig3}
\end{figure}


\subsection{Uncertainty Quantification}
\label{sec:uq_capabilities}

In this section, we examine our framework’s UQ capabilities across a range of SNRs. In addition, we compare the performance of our framework to the two most common approaches for data-driven AMC with UQ: a standalone CNN classifier \cite{ref29,ref33,ref35} and an AMC-based BNN \cite{ref2}. The CNN is a single deep learning classifier that provides state-of-the-art performance but limited UQ capabilities. BNNs are another deep learning-based AMC framework capable of quantifying uncertainty better than standalone deep learning models by generating predictive distributions over point estimates, similar to our proposed ensemble, during inference. Yet, as we will show in our empirical analysis, BNNs suffer in baseline classification performance in comparison to state-of-the-art standalone deep learning classifiers. Furthermore, we introduce an extra baseline: An SNR-aware weighted ensemble inspired by \cite{ref_weighted}, where we use an SNR-aware approach to train each model in an ensemble on a specific SNR and then calculate the weight of each model using the Shannon entropy. This approach introduces a baseline in which each model in the ensemble is trained on a unique SNR, thus distinguishing between epistemic uncertainty (model uncertainty) and aleatoric uncertainty (channel noise) in AMC tasks. Here, models with high entropy (e.g., classifiers trained on low SNR signals) are given a lower weight and models with low entropy (e.g., classifiers trained on high SNR signals) are given more weight (proportional to the Shannon entropy of each classifier). Note that although \cite{ref_weighted} does not explicitly apply a weighted ensemble to AMC, it shows the potential of a weighted ensemble to characterize aleatoric uncertainty so we adapt it for AMC as an additional baseline. Contrary to all of these methods, we will see that our proposed deep ensemble framework for AMC provides both state-of-the-art classification performance as well as robust UQ. 

Fig. \ref{fig2} shows the performance of our approach in comparison to each considered baselines in terms of our considered scoring metrics. As shown in Fig. \ref{fig2}, our ensemble-based approach consistently outperforms a standalone CNN, the BNN, and the weighted ensemble on both considered datasets. Notably, while the single CNN, the weighted ensemble, and our proposed ensemble exhibit relatively stable accuracy in both datasets, the BNN’s accuracy declines sharply on the RML 2018.01a dataset -- likely due to the more complex real-world signals in it compared to the GNU radio generated data in the RML 2016.10a dataset, highlighting its limitations in AMC applications. Despite these challenges, our proposed ensemble maintains the highest overall accuracy. Similarly, Fig. \ref{fig2} shows that our proposed ensemble also outperforms the standalone CNN, the BNN, and the weighted ensemble in terms of NLL, consistently achieving a lower NLL score across the entire SNR range. The BNN, in particular, consistently attains a high NLL, relative to our proposed ensemble, in both considered datasets at each considered SNR. Lastly, we see from Fig. \ref{fig2} that our ensemble consistently achieves the lowest Brier scores in all considered environments, further demonstrating its robustness relative to the other considered baselines. 

In Fig. \ref{fig:fig3}, we show the CI widths of low, medium, and high SNR values of our proposed ensemble model in comparison to BNNs and the weighted ensemble on correctly predicted signals. In this scenario, we omit results from the standalone CNN, as they produce point predictions, resulting in CI widths of zero. From Fig. \ref{fig:fig3}, we observe that, at $-10$ dB (low SNR), the BNN exhibits a mix of narrow and moderate CI widths, suggesting some level of uncertainty in low-SNR conditions. However, as the SNR increases to $0$ dB and $8$ dB, the BNN’s CI widths shift toward smaller values, indicating greater confidence in its predictions. This behavior aligns with expectations, as higher SNR leads to cleaner input data, reducing uncertainty. The weighted ensemble, on the other hand, attains narrower widths at low SNR, where higher uncertainty expression is more crucial, and higher widths at high SNR, where higher uncertainty expression is less crucial. In contrast, our proposed ensemble model maintains a higher degree of uncertainty across all SNR levels, with wider CI widths than both considered baselines. Moreover, as SNR increases, the ensemble also becomes more confident in its correct predictions, while still preserving some uncertainty in certain cases. Notably, our proposed ensemble achieves higher classification accuracy than the BNN and the weighted ensemble baseline (as shown in Fig. \ref{fig2}), despite its broader confidence intervals. This suggests that while the ensemble model expresses more uncertainty, it does not come at the cost of accuracy. Instead, it offers a more calibrated representation of confidence, ensuring that even in high-SNR conditions, some uncertainty is retained where appropriate.


\begin{figure}[t]
    \centering
    \subfloat{%
        \includegraphics[width=0.33\linewidth, height=0.24\linewidth]{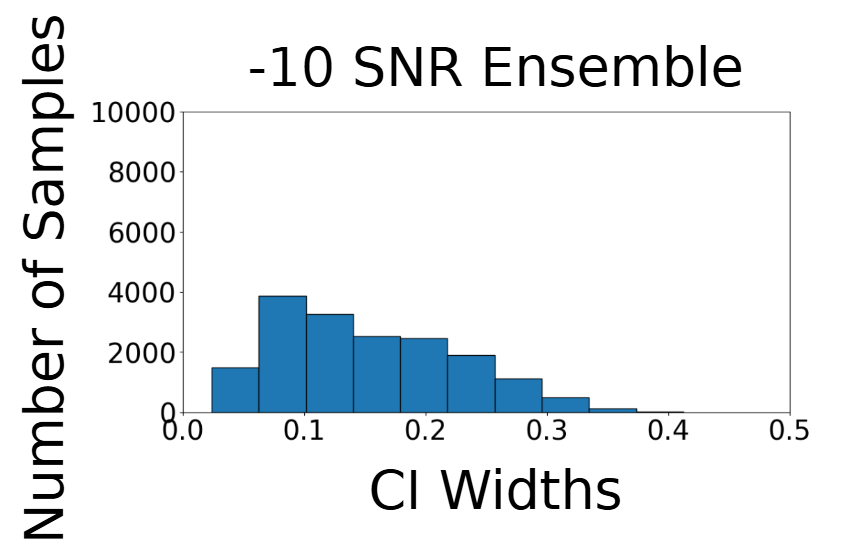}
        \label{fig:-10SNR_ensemble_inC}
    }
    \subfloat{%
        \includegraphics[width=0.33\linewidth, height=0.24\linewidth]{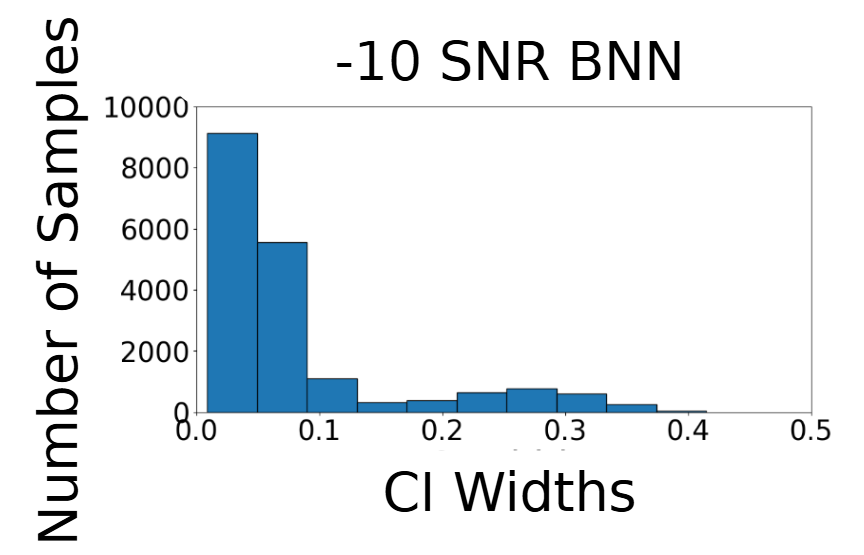}
        \label{fig:-10SNR_BNN_inC}
    }
    \subfloat{%
        \includegraphics[width=0.33\linewidth, height=0.24\linewidth]{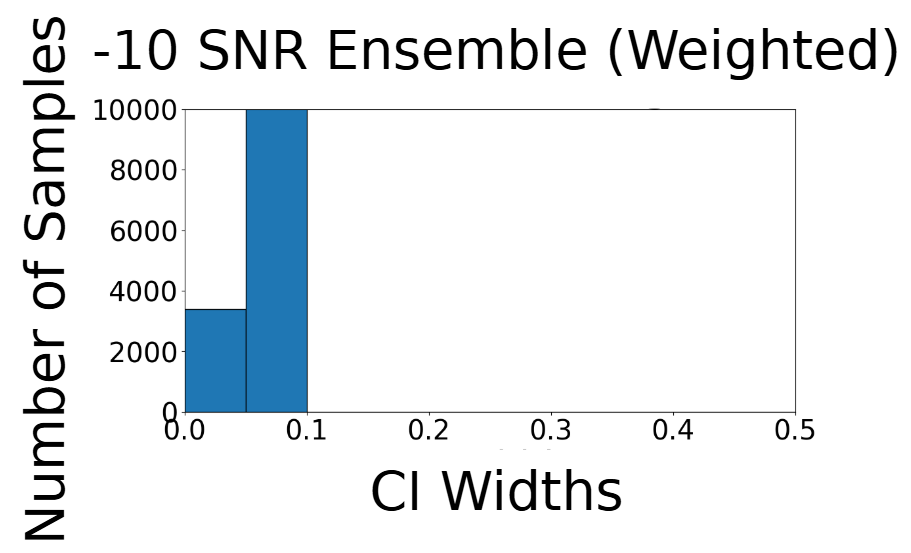}
        \label{fig:-10SNR_WEN_inC}
    }
    
    \subfloat{%
    \includegraphics[width=0.34\linewidth,height=0.24\linewidth]{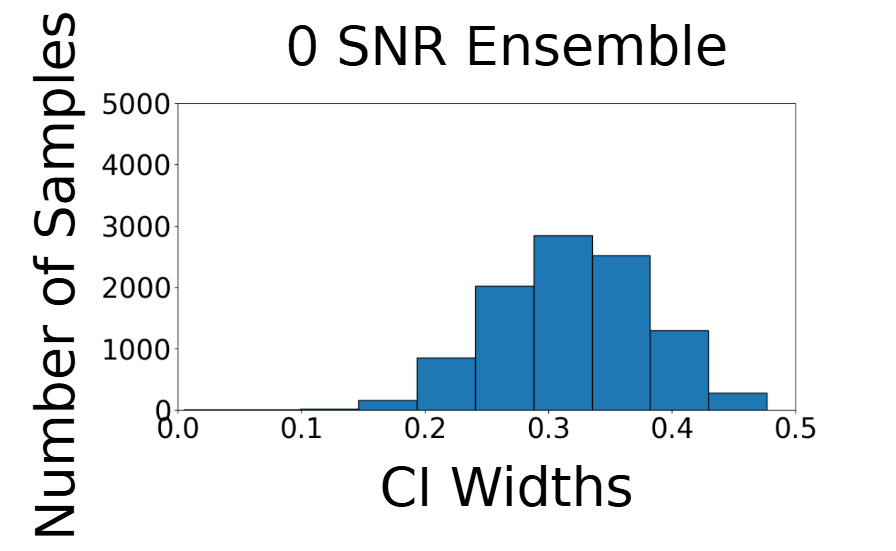}
        \label{fig:0SNR_ensemble_inC}
    }
    \subfloat{%
        \includegraphics[width=0.33\linewidth,height=0.24\linewidth]{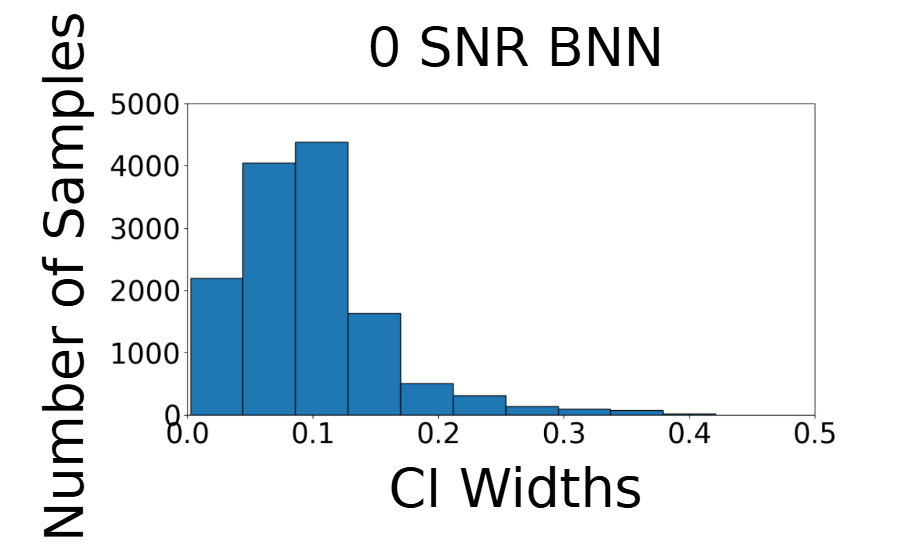}
        \label{fig:0SNR_BNN_inC}
    }
    \subfloat{%
        \includegraphics[width=0.33\linewidth, height=0.24\linewidth]{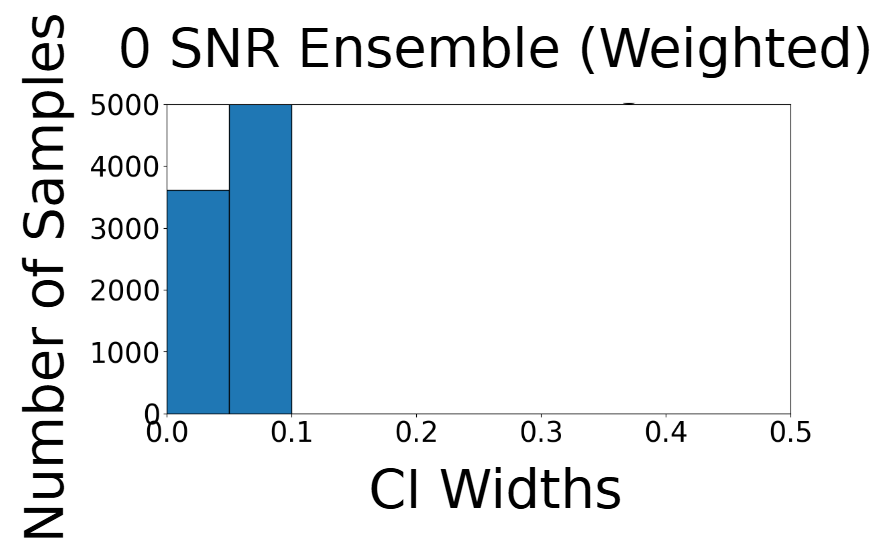}
        \label{fig:0SNR_WEN_inC}
    }

    \subfloat{%
        \includegraphics[width=0.33\linewidth,height=0.24\linewidth]{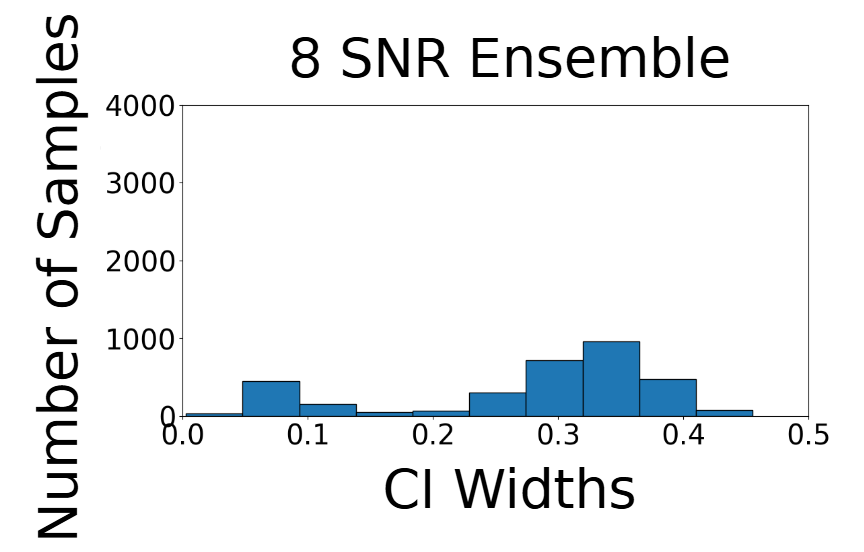}
        \label{fig:8SNR_ensemble_inC}
    }
    \subfloat{%
        \includegraphics[width=0.33\linewidth,height=0.24\linewidth]{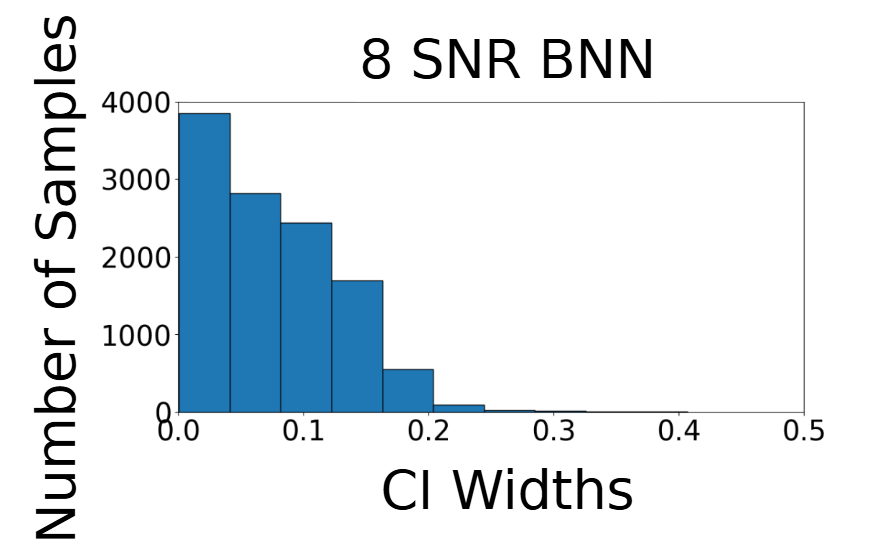}
        \label{fig:8SNR_BNN_inC}
    }
    \subfloat{%
        \includegraphics[width=0.33\linewidth, height=0.24\linewidth]{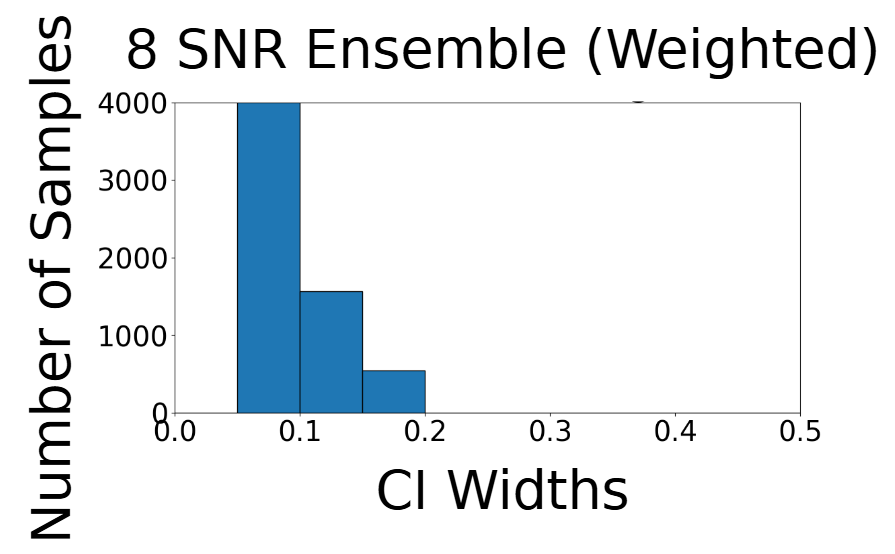}
        \label{fig:8SNR_WEN_inC}
    }

    \caption{
    Incorrectly predicted CI widths of our proposed ensemble (left column), the baseline BNN approach (middle column), and the baseline weighted ensemble (right column) on RML 2018.01a. Similar to Fig. \ref{fig:fig3}, we see here that our ensemble produces wider CI widths compared to both baselines, demonstrating that our proposed ensemble can more effectively characterize uncertainty estimates, particularly on incorrectly predicted signals.}
    
    \label{fig:fig4}
\end{figure}


We now turn our focus to the CI widths of low, medium, and high SNR values of our proposed ensemble model on incorrectly predicted signals. In this scenario, robust uncertainty is demonstrated with high CI widths as we would not want a model to express high confidence in a wrong AMC prediction. As shown in Fig. ~\ref{fig:fig4}, at $-10$ dB, the BNN and the weighted ensemble baseline primarily produce near-zero CI widths, meaning they are overconfident even when they are incorrect. This misplaced confidence persists at higher SNRs (e.g., $0$ dB and $8$ dB), indicating that both baselines fail to appropriately capture uncertainty in their misclassifications. In contrast, our proposed ensemble exhibits wider CI widths for incorrect predictions, suggesting that it remains more cautious, especially in low-SNR environments, where the performance of deep learning-based AMC classifiers is known to struggle. As the SNR increases, our proposed ensemble’s average CI width remains wide, reflecting an appropriate level of uncertainty even in moderate to low noise conditions. Meanwhile, the overconfidence of the other baselines remains evident across all SNR levels, failing to distinguish between correct and incorrect predictions in terms of uncertainty representation.

\begin{figure}[t]
    \centering
    \subfloat{%
        \includegraphics[width=0.48\linewidth]{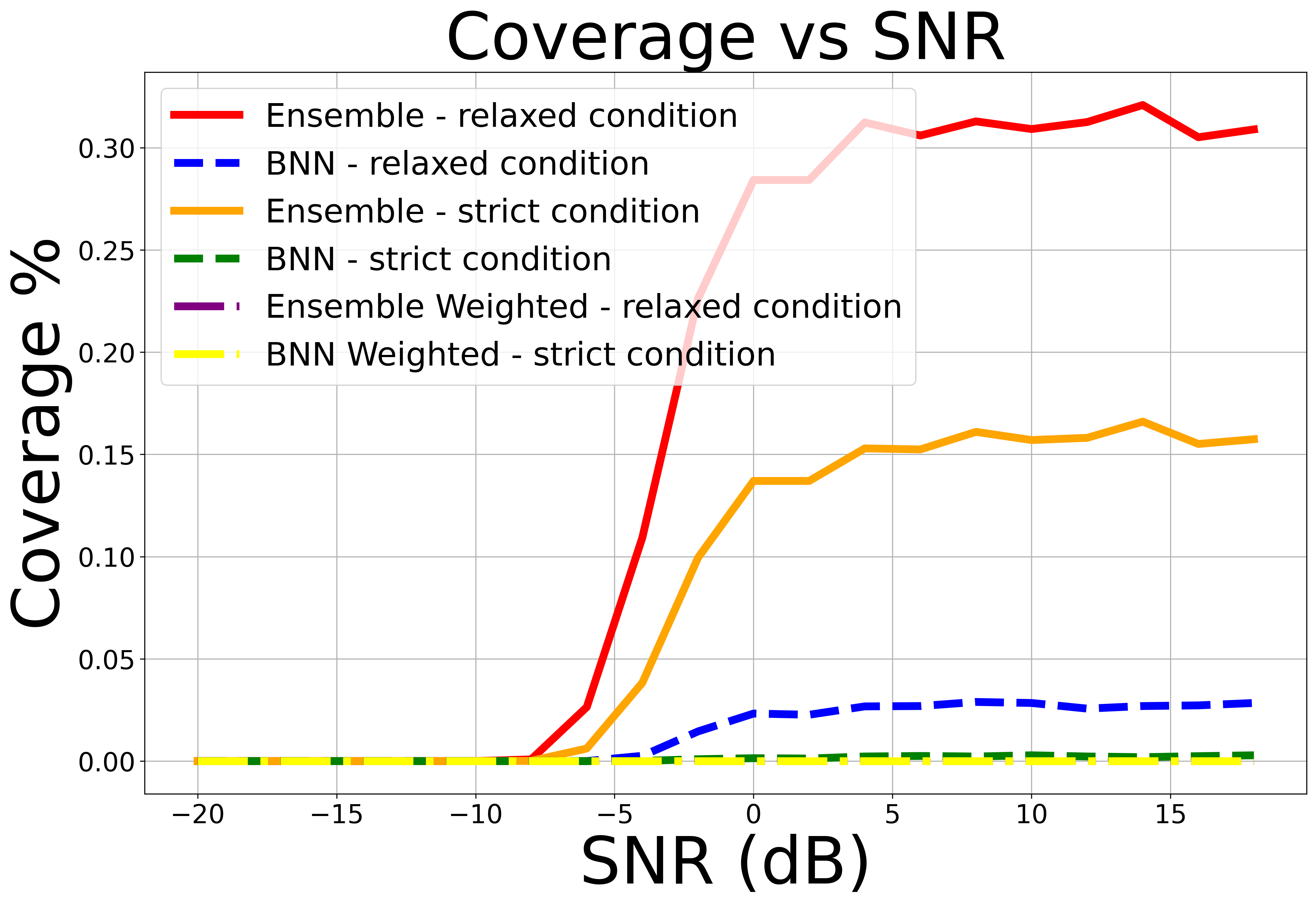}
        \label{fig:2016_coverage}
    }
    \subfloat{%
        \includegraphics[width=0.48\linewidth]{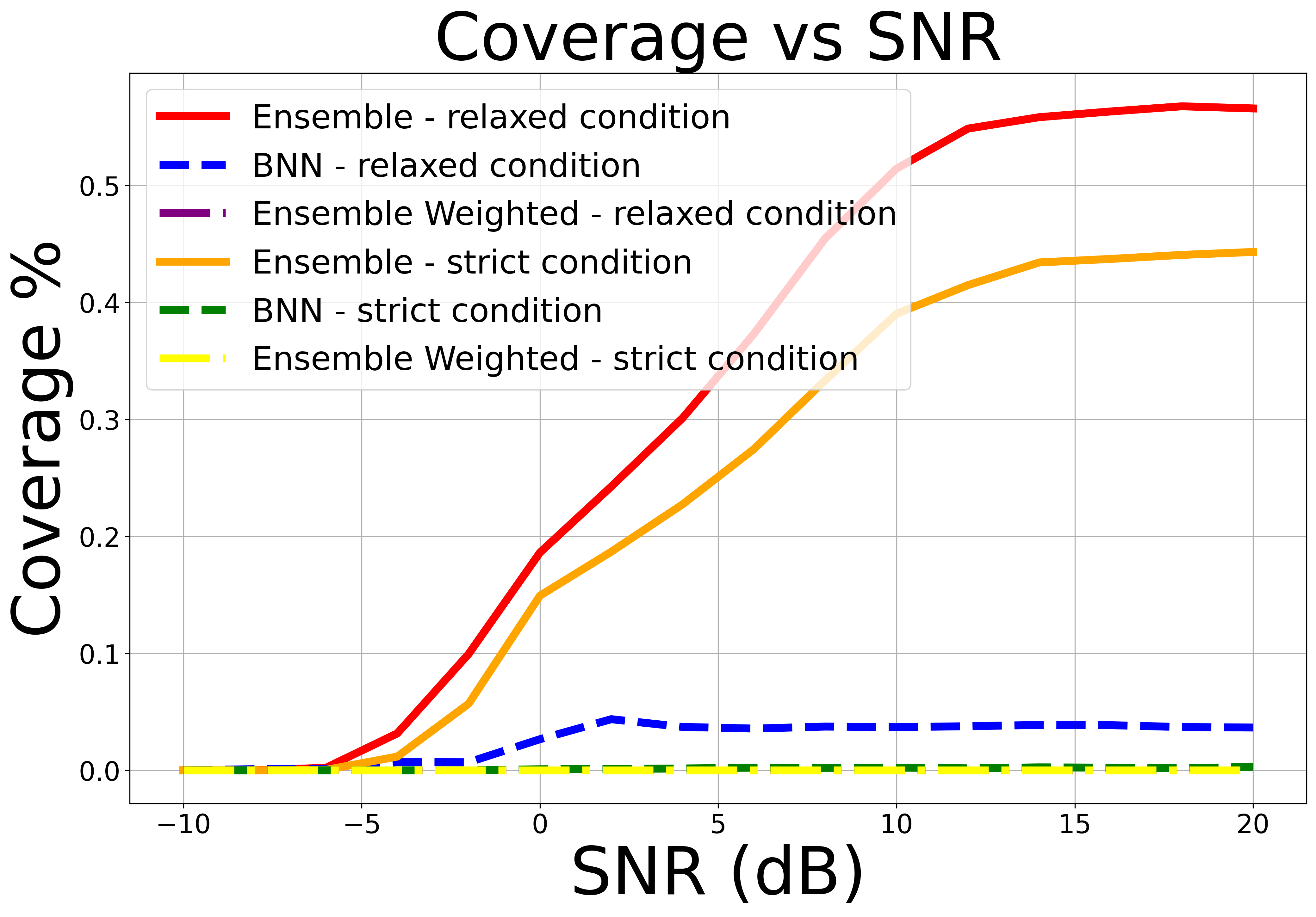}
        \label{fig:2018_coverage}
    }

    \caption{
    Coverage proportion, under strict and relaxed conditions, of our approach in comparison to each considered baseline. Here, we see that our proposed ensemble achieves a higher coverage proportion under both the strict and relaxed conditions for RML 2016.10a (left) and RML 2018.01a (right). This indicates that our proposed ensemble excels in its predictive interval, ensuring that the true class is more likely to be contained.}
    
    \label{fig:fig5}
\end{figure}

Next, we assess coverage proportions in Fig. \ref{fig:fig5}, which describe how well each model’s predictive intervals capture the true class. A higher coverage indicates stronger uncertainty representation. Here, we omit the coverage of the single model since it produces point estimates, which results in coverage proportions of $0$. In Fig. \ref{fig:fig5}, we see that the ensemble outperforms both considered baselines, particularly surpassing the coverage proportions of BNNs. Moreover, we see this trend hold for both the strict and relaxed definitions of coverage (as defined in Sec. \ref{sec:estimation_metrics}), indicating that our proposed ensemble can represent, to a higher degree than each considered baseline, both its uncertainty associated with its predicted class as well as its uncertainty associated with its prediction that the input does not belong to any other class.

\begin{figure}[t]
    \centering
    
    \subfloat{%
        \includegraphics[width=0.48\linewidth]{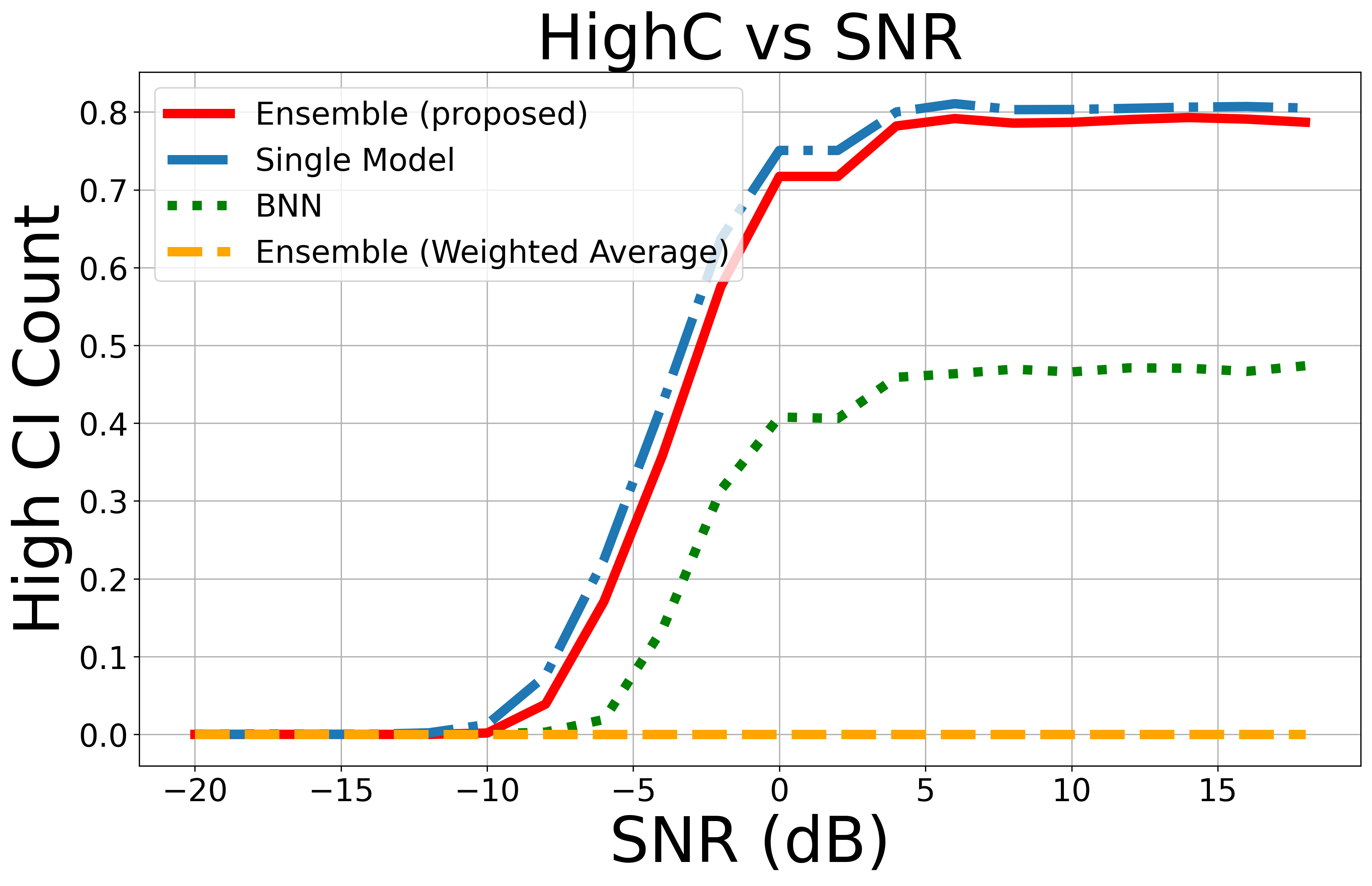}
        \label{fig:2016_highC}
    }
    \subfloat{%
        \includegraphics[width=0.48\linewidth]{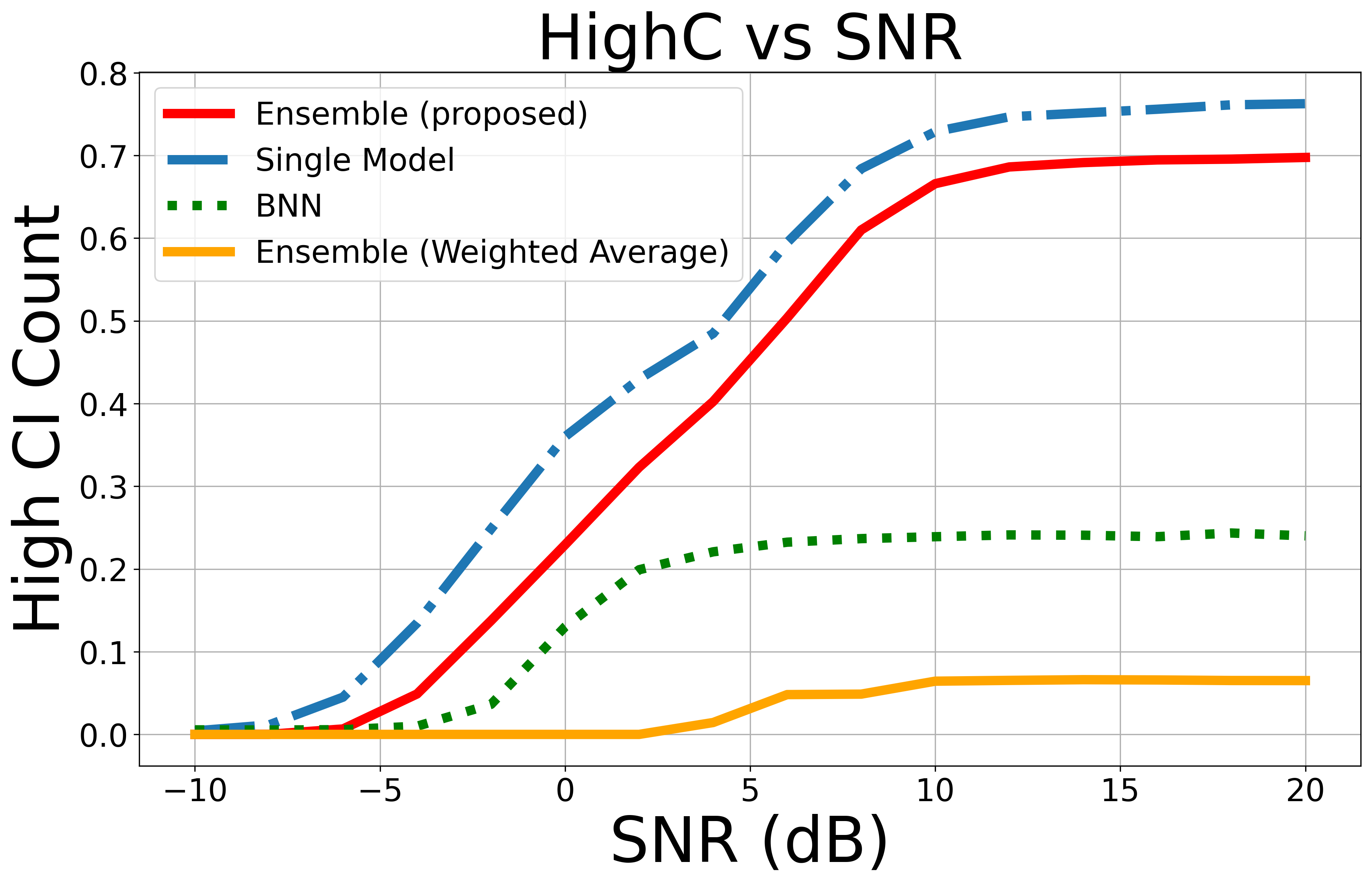}
        \label{fig:2018_highC}
    }

    \caption{High confidence proportion of our proposed ensemble in comparison to each considered baseline. In both RML 2016.10a (left) and RML 2018.01a (right), our ensemble maintains a middle level between the CNN, baseline BNN, and baseline weighted ensemble indicating that our proposed ensemble is neither overconfident nor underconfident and able to strike a UQ balance compared to the other models.}
    \label{fig:fig6}
\end{figure}

We now evaluate the high-confidence sets in Fig.~\ref{fig:fig6}, where we label a prediction as high-confidence if the model assigns at least an 80\% probability to a particular class (as elaborated on in Sec. \ref{sec:estimation_metrics}). The standalone CNN leads in this category, reflecting its tendency to be overconfident. Conversely, both the BNN and the weighted ensemble baselines struggle with underconfidence, consistently producing fewer high-confidence predictions. Our proposed ensemble model strikes a balanced approach, achieving strong accuracy while also retaining suitable uncertainty, even at higher SNR levels for its high-confidence prediction, reinforcing its ability to characterize uncertainty to a higher degree in comparison to the considered baselines.

\begin{figure}[t]
    \centering

    \subfloat{%
        \includegraphics[width=0.48\linewidth]{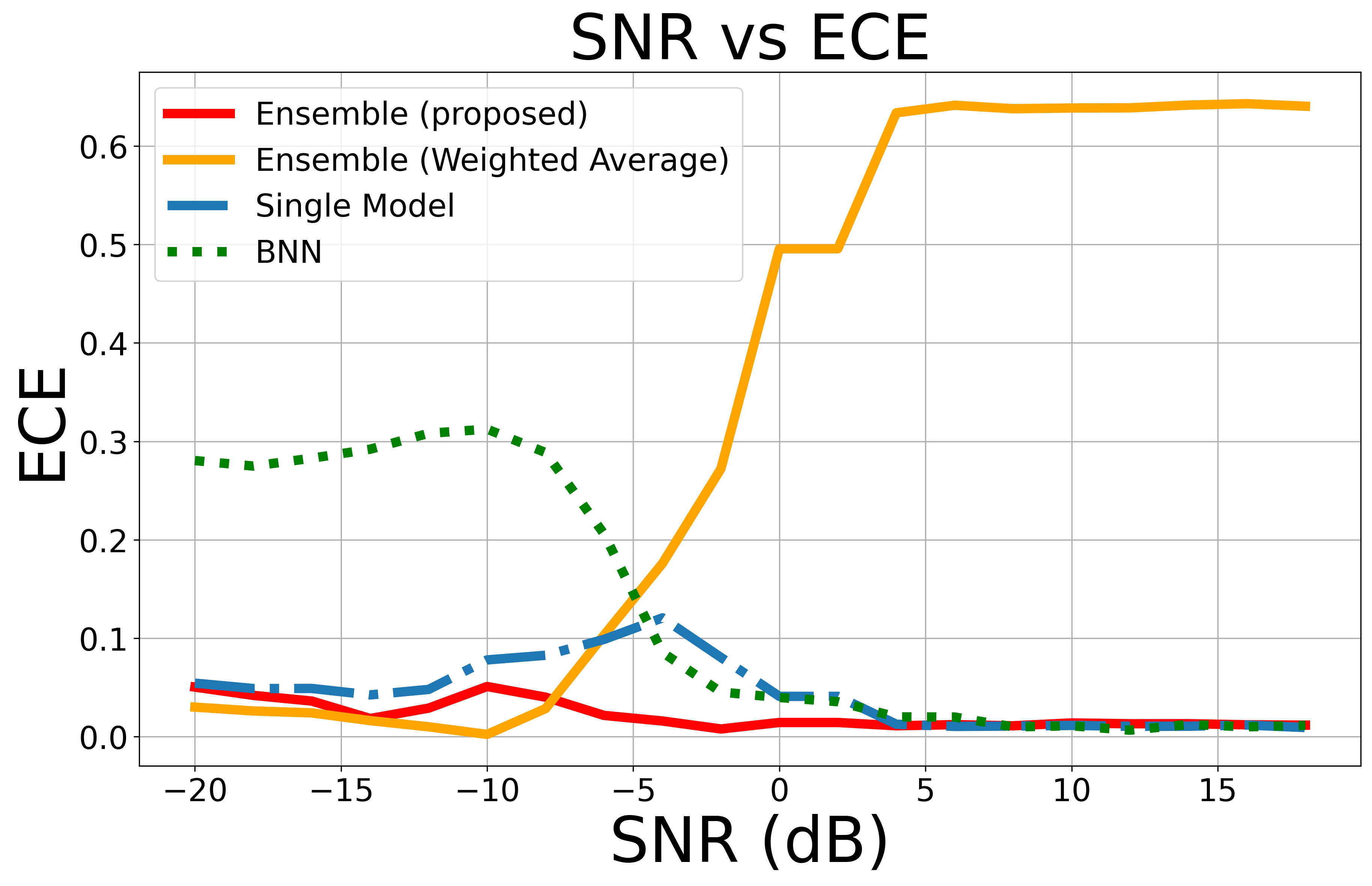}
        \label{fig:2016 ece}
    }
    \subfloat{%
        \includegraphics[width=0.48\linewidth]{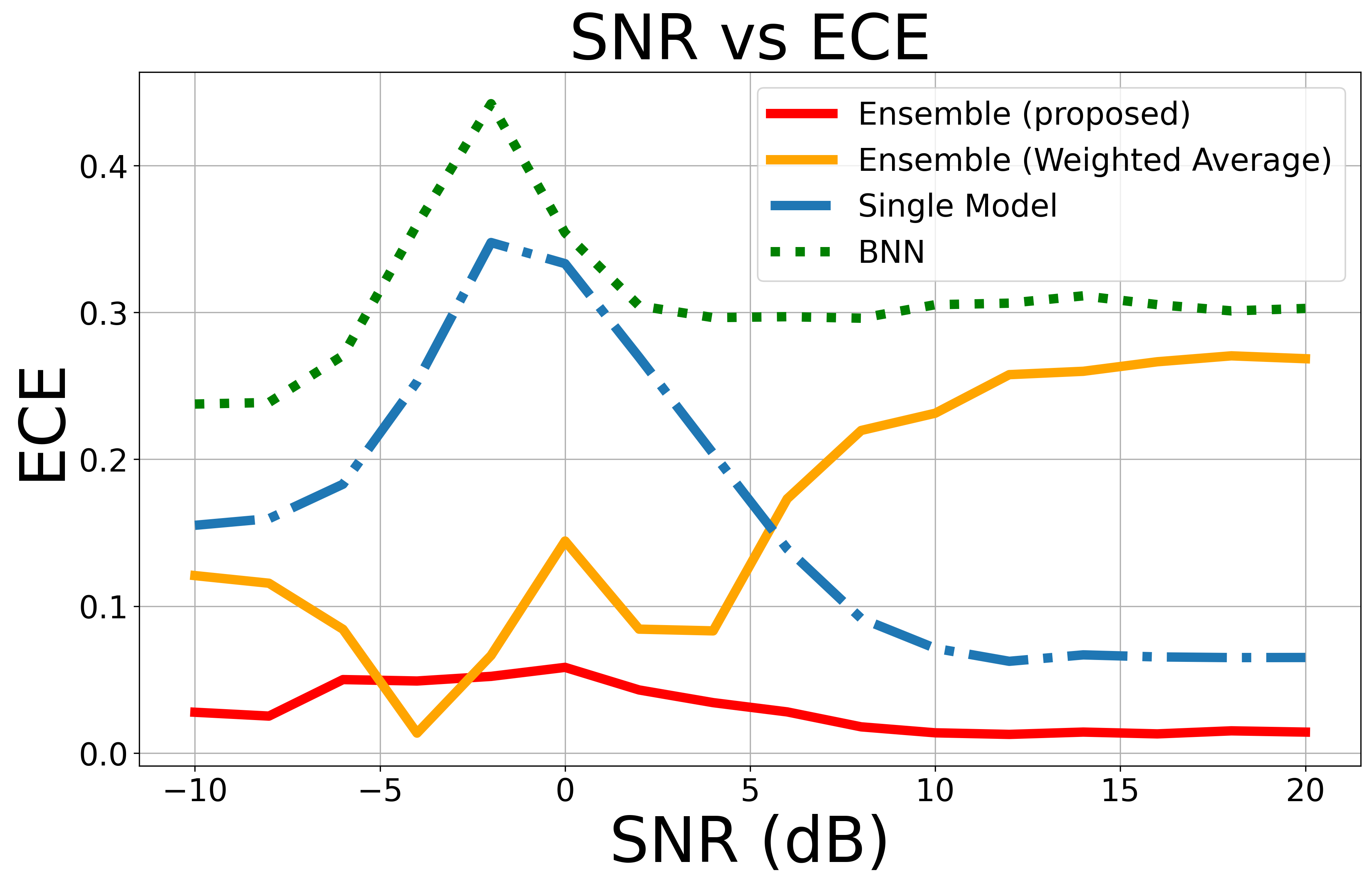}
        \label{fig:2018_ECE}
    }
    
    \subfloat{%
        \includegraphics[width=0.48\linewidth]{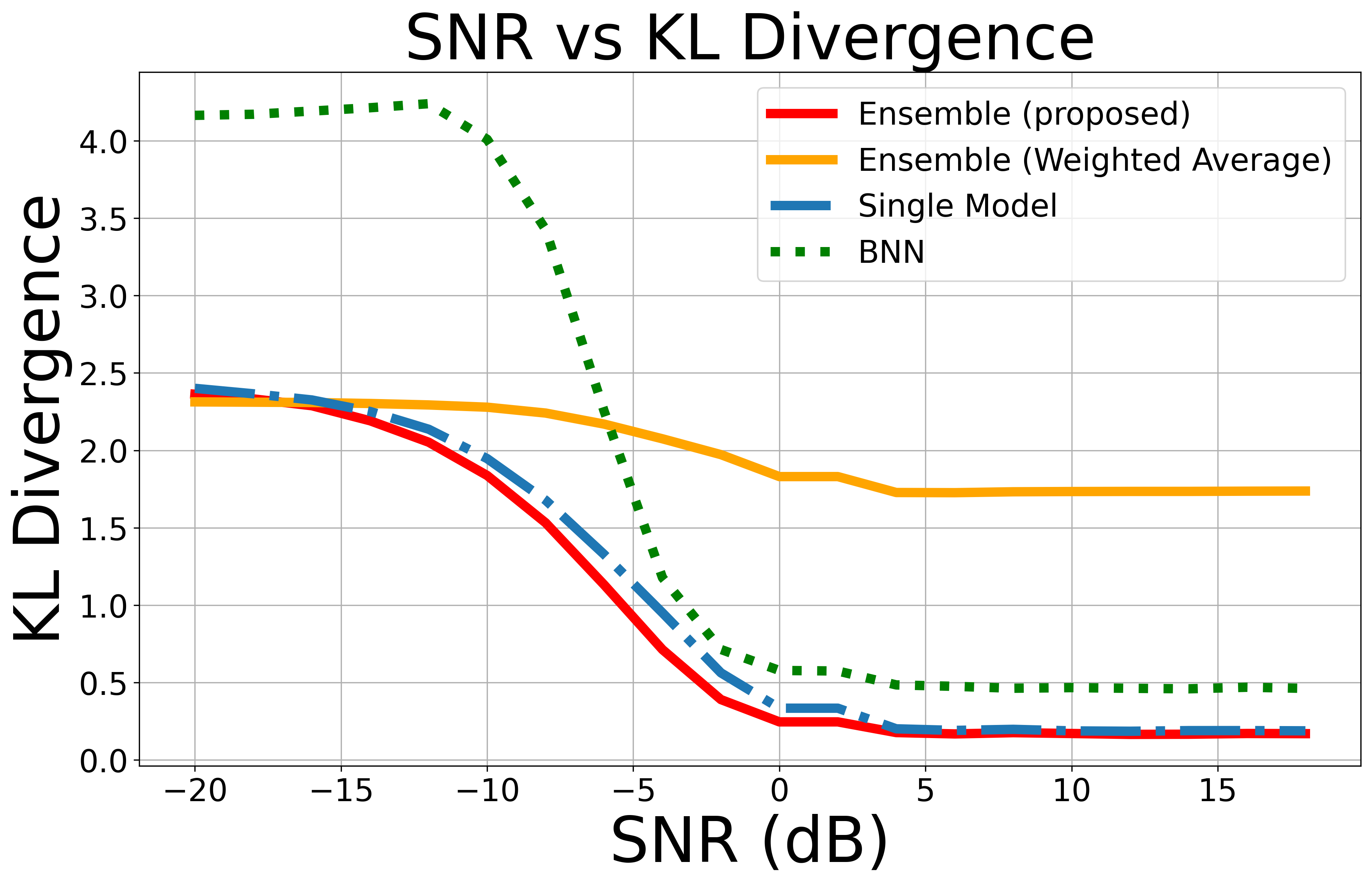}
        \label{fig:2016 kl}
    }
    \subfloat{%
        \includegraphics[width=0.48\linewidth]{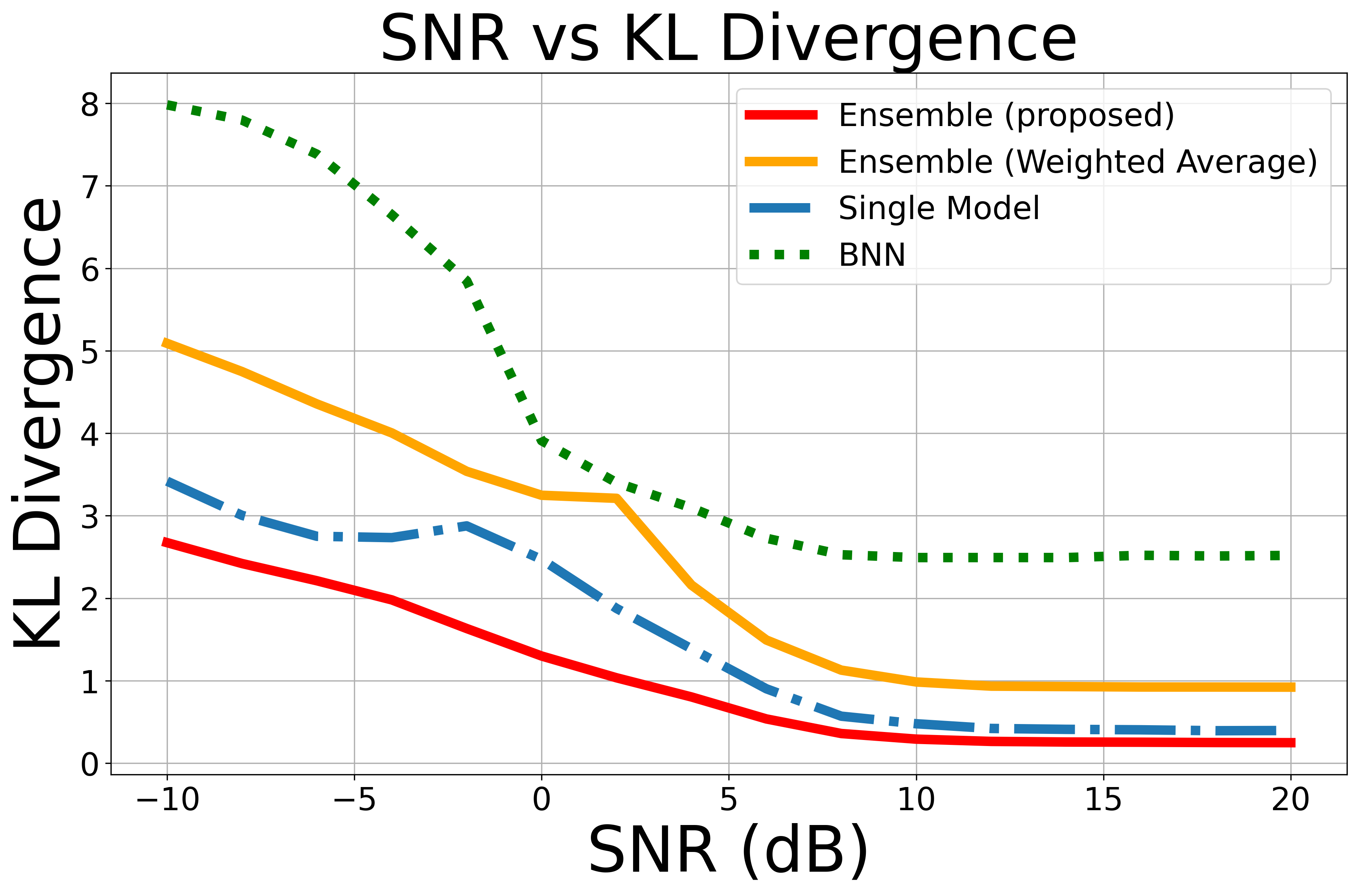}
        \label{fig:2018 kl}
    }

    \caption{
   ECE and KL Divergence of our proposed ensemble in comparison to each considered baseline on the 2016.10a (left) and 2018.01a (right) dataset. Here, we see that our proposed ensemble is able to achieve the lowest ECE and KL divergence compared to the baseline showing improved calibration.
    }
    \label{fig:fig7}
\end{figure}

Finally, we analyze the UQ capabilities of our framework by measuring the ECE and KL Divergence. A lower ECE indicates better model calibration, and as shown in Fig.~\ref{fig:fig7}, our ensemble consistently achieves the smallest calibration error across all SNR levels, whereas the BNN spikes and trails behind. Similarly, the weighted ensemble baseline exhibits a higher degree of ECE across the entire SNR range, indicating its confidence estimates are not well-calibrated to its accuracy. The same trend holds for the KL divergence, where our ensemble maintains the lowest divergence from the true predictive distribution, outperforming both the standalone CNN, the BNN, and the weighted ensemble baselines showcasing more expressive and robust UQ capabilities.

In the context of AMC, it is useful to distinguish between epistemic and aleatoric uncertainty. Epistemic uncertainty arises from model limitations (e.g., insufficient training data) and can potentially be reduced by larger training sets or higher-parameter models. Aleatoric uncertainty, on the other hand, stems from inherent noise in the received signals (e.g., low SNR or OOD data) and is a result of the quality of the received signals used for training. Our results shown in Figs. \ref{fig2} -- \ref{fig:fig7} demonstrate that our proposed ensemble-based framework captures both types of uncertainty: the diversity among independently trained models reflects epistemic uncertainty, while the variability in predictions across signal conditions (e.g., at low SNRs) captures aleatoric uncertainty. Compared to the BNN and weighted ensemble baselines, our method more effectively disentangles these sources, as shown by its higher confidence interval widths on misclassified samples (epistemic) and its improved calibration under noisy conditions (aleatoric).


\subsection{Out-of-Distribution Performance}
\label{sec:ood_performance}

Here, we examine how our framework handles OOD scenarios by examining its effects on adversarial examples. Although they are typically below the noise floor of wireless signals, adversarial perturbations can significantly degrade the performance of neural network–based AMC methods by inducing models to output incorrect predictions due to their underlying shifted distributions \cite{ref54}. 

\begin{figure}[t]
    \centering
    \subfloat{%
        \includegraphics[width=0.48\linewidth]{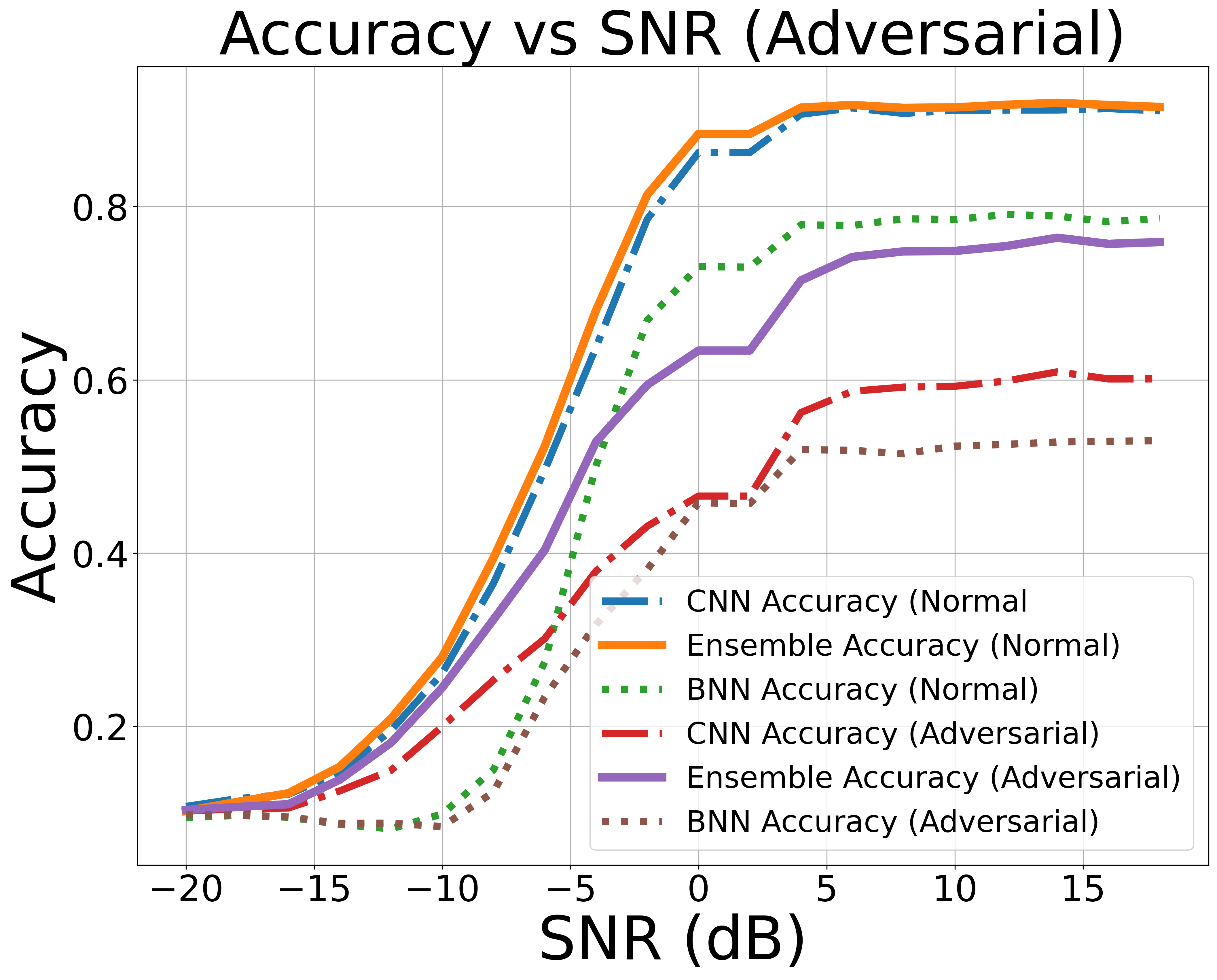}
        \label{fig:2016_adv_acc}
    }
    \subfloat{%
        \includegraphics[width=0.48\linewidth]{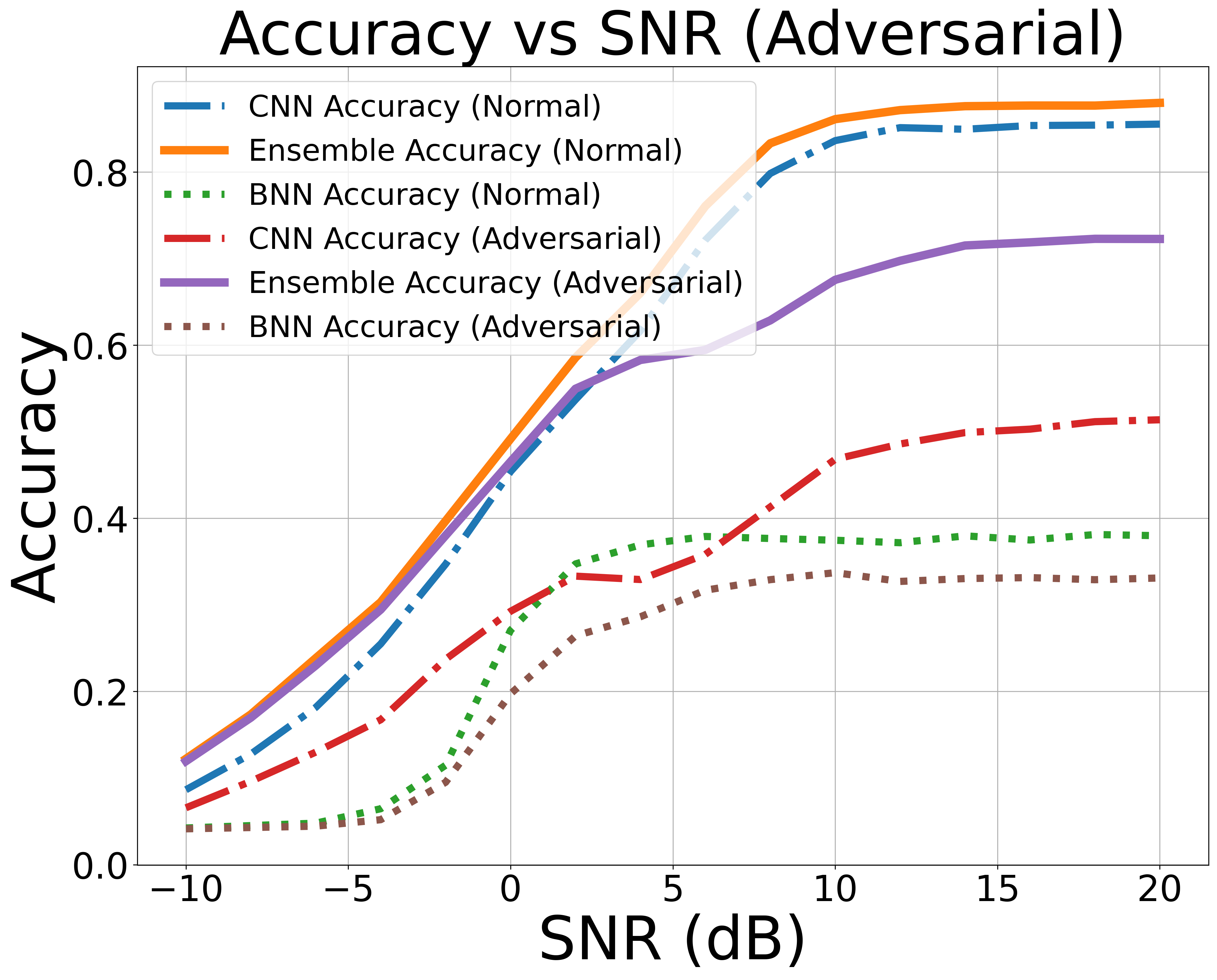}
        \label{fig:2018_adv_acc}
    }

    \caption{
    Accuracy after applying a constant perturbation of $5$ dB across varying SNRs on 2016.10a (left) and 2018.01a (right). The results demonstrate that, under both normal and adversarial conditions, our ensemble model achieves the best performance at high and low SNR values. 
    }
    \label{fig:fig8}
\end{figure}

\begin{figure}[t]
    \centering
    \subfloat{%
        \includegraphics[width=0.48\linewidth]{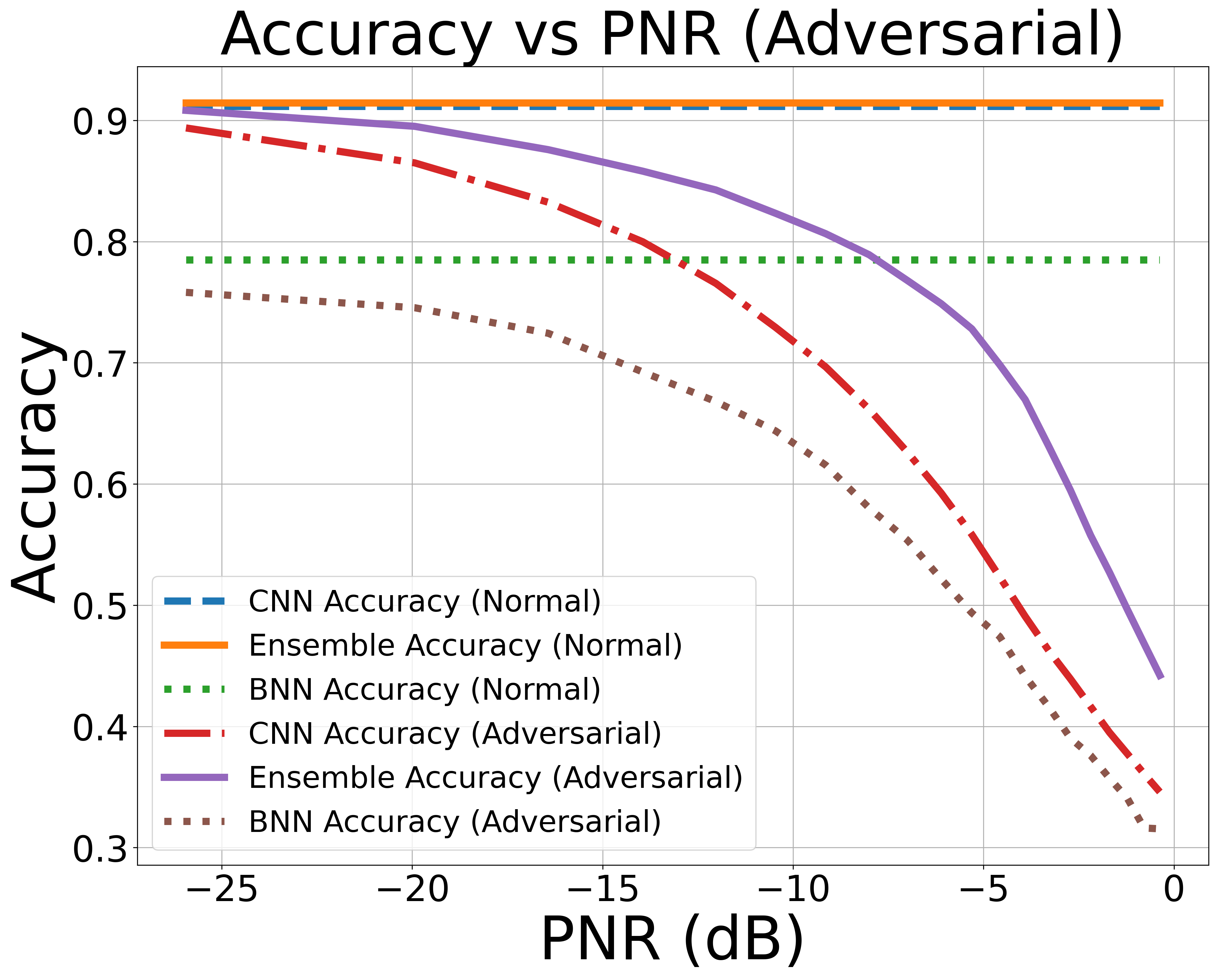}
        \label{fig:2016_adv_accPNR}
    }
    \subfloat{%
        \includegraphics[width=0.48\linewidth]{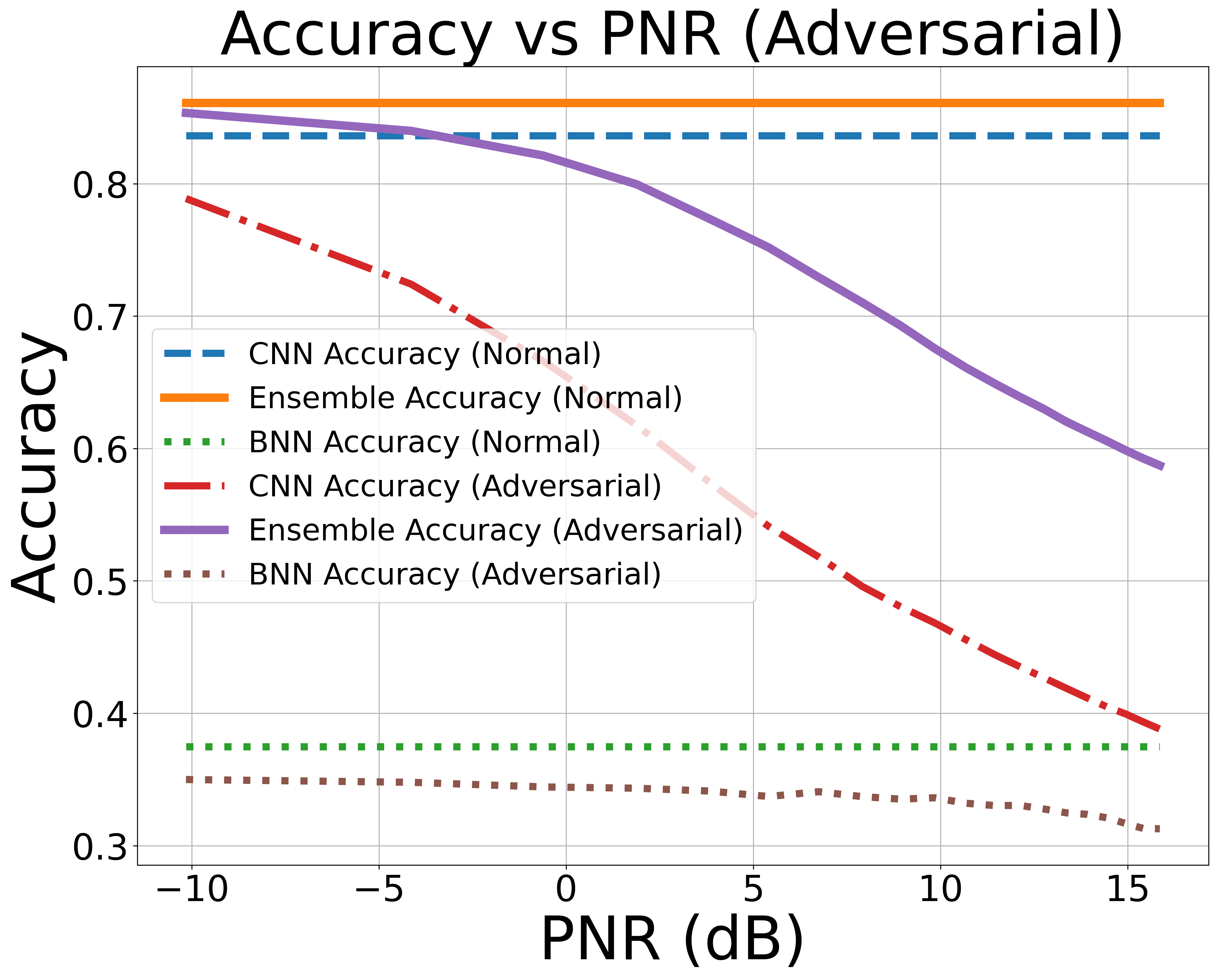}
        \label{fig:2018_adv_accPNR}
    }

    \caption{
    Accuracy after varying the perturbation on $10$ dB SNR signals on the 2016.10a (left) and 2018.01a (right) dataset. Here, we see that our proposed ensemble is able to maintain higher performance in comparison to the considered baselines as the PNR increases and approaches the noise floor. 
    }
    \label{fig:fig9}
\end{figure}

We first examine the effect of adding a constant PNR of 5 dB across the entire SNR range. Fig.~\ref{fig:fig8} displays each model’s accuracy on unperturbed samples for comparison. While the ensemble and the standalone CNN show similar performance on clean signals, the ensemble significantly outperforms the CNN on OOD AMC signals. This disparity highlights the ensemble’s superior UQ capabilities in challenging signal environments. We next examine our proposed ensemble's accuracy across increasing PNR at a constant SNR of $10$ dB. As the PNR increases, the perturbed sample is shifted further from its unperturbed counterpart, thereby increasing the chance of misclassification and further shifting the distribution of the adversarial examples. Fig. \ref{fig:fig9} includes accuracies under normal (unperturbed) conditions, shown as straight lines since SNR remains constant. Here, we see that even as the PNR increases, our proposed ensemble is able to withstand the performance degradation to a greater extent than both the CNN and the BNN. Overall, these findings highlight our proposed ensemble’s ability to retain robust performance under adversarial attacks, reinforcing its ability to characterize UQ to a greater extent in comparison to the considered baselines while simultaneously striking a balance between high confidence and well-calibrated uncertainty.


\subsection{Discussion}
\label{sec:complexity analysis}

Here, we examine why our proposed ensemble outperforms BNNs, which are specifically designed to deliver robust UQ. While BNNs theoretically offer a Bayesian measure of uncertainty via their posterior distribution-based estimates, they often require unrealistic approximations that limit their ability to attain a high baseline accuracy in AMC tasks. For example, BNNs use variational inference, which by definition assumes independence between weights in the model. However, in AMC, the model operates on raw time-domain IQ samples, which are highly correlated with their nearby features. As a result, BNNs become limited from achieving state-of-the-art accuracy on AMC data as the correlation between features are highly indicative in discerning modulation type and BNNs are unable to exploit such correlations for high accuracy. In contrast, CNNs are not limited by the assumption of feature independence and, thus, ensembles of independently trained CNNs are able to exploit spatial regions of the received waveform (i.e., various portions of the time-domain signal) and are able to learn distinctive features between signals, resulting in robust uncertainty estimates through implicit posterior approximations. This makes our proposed ensemble particularly effective in comparison to BNNs, despite lacking a fully Bayesian foundation.

Furthermore, BNNs assume Gaussian priors on their weights (often with standard zero-mean, unit-variance distributions), which tends to break down for AMC as each feature in a received waveform is not necessarily Gaussian in real-world radio data, and moreover, AMC data may require more accurate assumptions on their priors to achieve high accuracy. For example, the channel distribution of received signals could differ from priors, hindering the ability of BNNs to effectively learn distinguishing features if an incorrect channel distribution assumption is made. CNNs, in contrast require no such prior assumption on the channel distribution of received signals and are not limited to learning weights of a certain distribution, allowing more expressive power and directly allowing higher AMC performance without specific channel assumptions. Thus, ensembles of CNNs, as proposed in our framework, not only achieve higher classification performance compared to BNNs, but they also express uncertainty more effectively due to their stronger expressive abilities.

Finally, as shown in Fig. \ref{fig2} -- Fig. \ref{fig:fig9}, our ensemble consistently outperforms the BNN baseline in both predictive accuracy and uncertainty calibration metrics (e.g., NLL and ECE). These results support the hypothesis that our proposed deep ensemble-based UQ approach offers more robust performance, even if BNNs offer a theoretically principled approach to UQ.

\section{Conclusion} \label{conclusion_sec}

Deep learning (DL) has been shown to provide cutting-edge performance in automatic modulation classification (AMC). However, DL-based AMC models often exhibit overconfidence in their predictions, with no associated measure of uncertainty. This issue is especially evident in low signal to noise ratio (SNR) conditions and when handling out-of-distribution (OOD) samples. In this work, we proposed a deep ensemble framework for AMC, which is capable of retaining the state-of-the-art performance of DL-based AMC classifiers while simultaneously providing robust uncertainty quantification (UQ) metrics. We demonstrated our ensemble's ability to achieve robust performance across multiple UQ metrics such as the negative log-likelihood, Brier score, prediction interval widths, prediction coverage, and high-confidence predictions. In comparison to standalone CNNs, weighted ensembles, and Bayesian Neural Networks (BNNs), we showed that our framework achieved better UQ estimates overall, particularly in low SNR and OOD environments. Moreover, our proposed framework is scalable to any DL architecture, allowing state-of-the-art performance to be extended to incorporate higher classification performance as well as uncertainty quantification on any DL-based AMC framework. Future work will explore the resilience of our approach in more versatile environments such as the distributed AMC scenario, which requires UQ in federated learning with varying channel conditions and adversarial interference at each receiver, and in environments with insufficient channel state information (CSI) knowledge.

\bibliography{references}

\bibliographystyle{IEEEtran}

\begin{IEEEbiography}
[{\includegraphics[width=1in,height=1.25in,clip,keepaspectratio,angle=90]{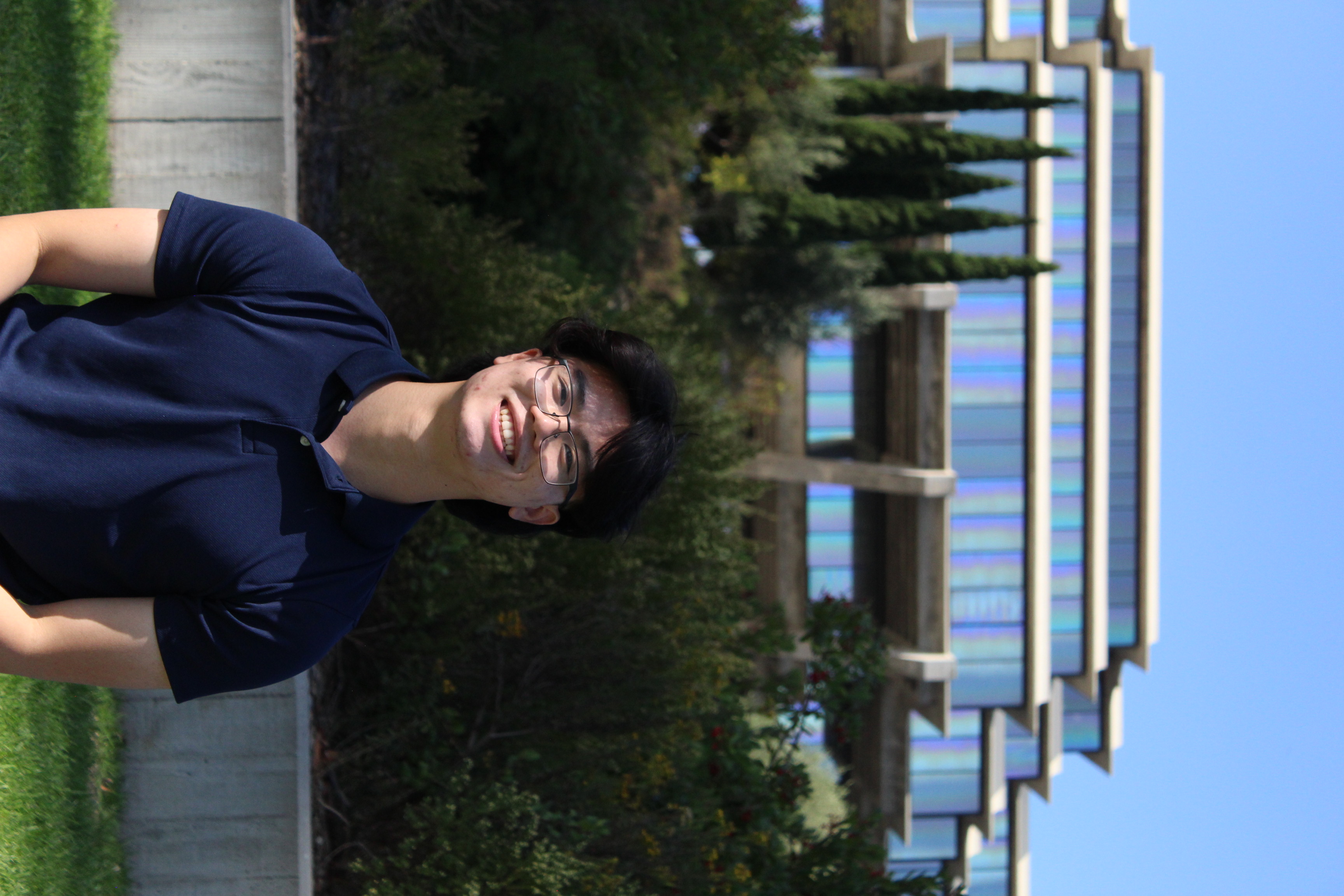}}]{Huian Yang} is currently pursuing his B.S. in Computer Engineering at the University of California San Diego. His research focuses on deep learning, with additional interests in embedded systems for robotics.
\end{IEEEbiography}

\begin{IEEEbiography}
[{\includegraphics[width=1in,height=1.25in,clip,keepaspectratio]{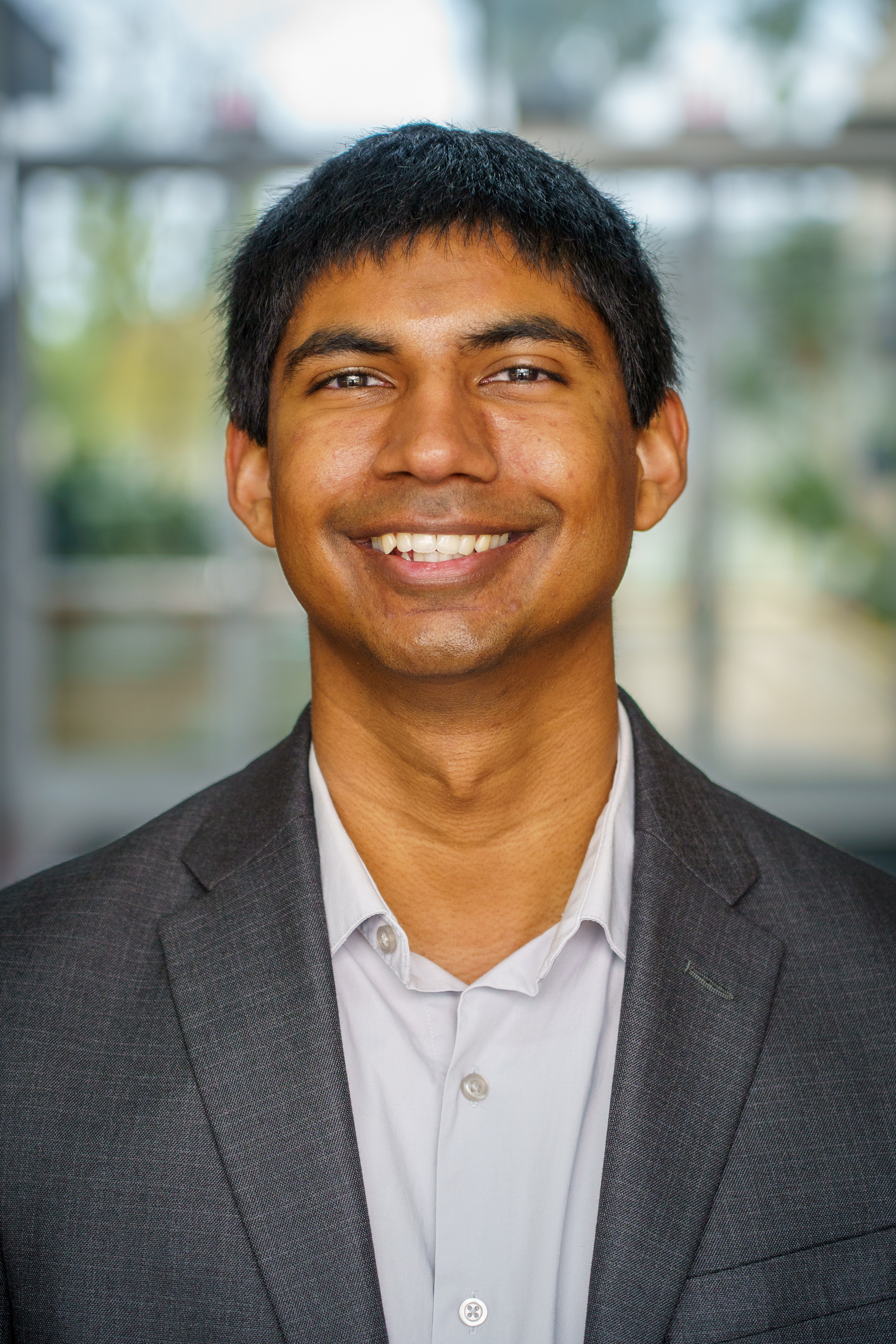}}]{Rajeev Sahay} received the B.S. degree in electrical engineering from The University of Utah, Salt Lake City, UT, USA, in 2018, and the M.S. and Ph.D. degrees in electrical and computer engineering from Purdue University, West Lafayette, IN, USA, in 2021 and 2022, respectively. Currently, he is a faculty member in the Department of Electrical and Computer Engineering at UC San Diego. He was the recipient of the Purdue Engineering Dean’s Teaching Fellowship and was named an Exemplary Reviewer by the IEEE Wireless Communications Letters. His research interests lie in the intersection of networking and machine learning, especially in their applications to wireless communications and engineering education. 
\end{IEEEbiography}

\end{document}